\documentclass[
superscriptaddress,
preprint,
amsmath,amssymb,
]{revtex4-2}

\usepackage{graphicx}
\usepackage{dcolumn}
\usepackage{changepage} 
\usepackage{lipsum}

\usepackage{bm}
\usepackage{subfigure}
\usepackage{booktabs}
\usepackage{graphicx}
\usepackage{upgreek}
\usepackage{subfigure}
\usepackage{color}
\usepackage{amsmath}
\usepackage[percent]{overpic}
\usepackage{makecell}
\usepackage{lineno}
\begin{document}
\title{Electrically pumped ultra-efficient quantum frequency conversion on thin film lithium niobate chip}
\author{Xina Wang}\thanks{These authors contributed equally to this work}
\affiliation{Hefei National Research Center for Physical Sciences at the Microscale and School of Physical Sciences,
University of Science and Technology of China, Hefei, Anhui 230026, China}
\affiliation{Jinan Institute of Quantum Technology and CAS Center for Excellence in Quantum Information and Quantum Physics, University of Science and Technology of China, Jinan, 250101, China}
\author{Xu-Feng Jiao}\thanks{These authors contributed equally to this work}
\affiliation{Hefei National Research Center for Physical Sciences at the Microscale and School of Physical Sciences,
University of Science and Technology of China, Hefei, Anhui 230026, China}
\affiliation{Jinan Institute of Quantum Technology and CAS Center for Excellence in Quantum Information and Quantum Physics, University of Science and Technology of China, Jinan, 250101, China}
\affiliation{Hefei National Laboratory, University of Science and Technology of China, Hefei, Anhui 230088, China}

\author{Bo Cao}
\affiliation{Jinan Institute of Quantum Technology and CAS Center for Excellence in Quantum Information and Quantum Physics, University of Science and Technology of China, Jinan, 250101, China}
\author{Yang Liu}
\affiliation{Jinan Institute of Quantum Technology and CAS Center for Excellence in Quantum Information and Quantum Physics, University of Science and Technology of China, Jinan, 250101, China}
\affiliation{Hefei National Laboratory, University of Science and Technology of China, Hefei, Anhui 230088, China}
\author{Xiu-Ping Xie}
\affiliation{Jinan Institute of Quantum Technology and CAS Center for Excellence in Quantum Information and Quantum Physics, University of Science and Technology of China, Jinan, 250101, China}
\affiliation{Hefei National Laboratory, University of Science and Technology of China, Hefei, Anhui 230088, China}
\author{Ming-Yang Zheng}
\affiliation{Jinan Institute of Quantum Technology and CAS Center for Excellence in Quantum Information and Quantum Physics, University of Science and Technology of China, Jinan, 250101, China}
\affiliation{Hefei National Laboratory, University of Science and Technology of China, Hefei, Anhui 230088, China}
\author{Qiang Zhang}
\affiliation{Hefei National Research Center for Physical Sciences at the Microscale and School of Physical Sciences,
University of Science and Technology of China, Hefei, Anhui 230026, China}
\affiliation{Jinan Institute of Quantum Technology and CAS Center for Excellence in Quantum Information and Quantum Physics, University of Science and Technology of China, Jinan, 250101, China}
\affiliation{Hefei National Laboratory, University of Science and Technology of China, Hefei, Anhui 230088, China}
\author{Jian-Wei Pan}
\affiliation{Hefei National Research Center for Physical Sciences at the Microscale and School of Physical Sciences,
University of Science and Technology of China, Hefei, Anhui 230026, China}
\affiliation{Hefei National Laboratory, University of Science and Technology of China, Hefei, Anhui 230088, China}

\date{\today}
\date{\today}


\maketitle
\begin{adjustwidth}{0.5cm}{0.5cm}

\textbf{Quantum frequency conversion (QFC) plays a crucial role in constructing seamless interconnection between quantum systems operating at different wavelengths. To advance future quantum technology, chip-scale integrated QFC components, featuring high efficiency, small footprint, low power consumption and high scalability, are indispensable. In this work, we demonstrate the first hybrid integrated QFC chip on thin film lithium niobate platform that connects the telecom and visible bands. Benefiting from the periodically poled microring resonator with ulta-high normalized conversion efficiency of 386,000 \%/W, an ultra-low pump power of 360 \(\mu\)W is achieved which is more than two orders of magnitude lower than traditional straight waveguide scheme. By injecting current into the chip, an on-chip quantum efficiency of 57\% and a noise count of \(\sim\) 7k counts per second are achieved. Such an electrically pumped, integrated and scalable QFC chip would significantly advancing the integration of quantum network and the development of chip-scale quantum optical systems.}
\end{adjustwidth}
\vspace{2em}

\section{Introduction}
Quantum frequency conversion (QFC), enabling coherent translation of photon frequencies while preserving quantum statistics, is a crucial technology for achieving wavelength compatibility in quantum science and technology~\cite{1,2}. This capability has been utilized to address critical frequency mismatches between different quantum systems, including quantum memories~\cite{15,4,3,8,14,5,van2020long,10}, single-photon sources~\cite{6,7,de2012quantum,singh2019quantum,Li:18}, fiber-based quantum network~\cite{12,13,9,11,van2022entangling,liu2024creation} and satellites- or unmanned aircraft-based free-space quantum communication system~\cite{16}, promoting the development of quantum network. Furthermore, QFC-based photon detection and interference have found widespread applications in diverse fields, including quantum key distribution~\cite{17}, single-photon lidar and imaging~\cite{18,wang2021non,wang2023mid}, and astrophysical interferometry~\cite{20,darre2016first,21,liu2021improved}.

Over the past decades, QFCs have mainly been achieved through sum-frequency generation (SFG) and difference-frequency generation (DFG) processes in bulk crystals and weakly confined waveguide structures~\cite{3,4,5,6,7,8,9,10,11,12,13,14,15,17,18,19,20,van2020long,van2022entangling,liu2024creation,darre2016first,21,liu2021improved,wang2023mid,Li:18,de2012quantum,singh2019quantum,wang2021non,PhysRevApplied.14.034035}. Recently, thin-film lithium niobate (TFLN) platforms have emerged as promising candidates for on-chip QFC implementations, leveraging their exceptional optical nonlinearity, flexible ferroelectric domain engineering, and strong optical confinement~\cite{25,26}. Nevertheless, current TFLN-based systems typically rely on optically pumped architectures requiring external bulky lasers, which imposes significant footprint constraints and scalability limitations~\cite{niu2020optimizing,27,29,28}. To overcome these challenges, electrically pumped configurations with integrated pump laser on TFLN chip offers a promising solution.

However, the typical output power of a laser chip is only a few milliwatts to tens of milliwatts, while traditional QFC requires a pump power at the level of several hundred milliwatts~\cite{3,4,5,6,7,8,9,10,11,12,13,14,15,17,18,19,20,van2020long,van2022entangling,liu2024creation,darre2016first,21,liu2021improved,wang2023mid,Li:18,de2012quantum,singh2019quantum,wang2021non,25,26,niu2020optimizing,27,29,PhysRevApplied.14.034035}. Furthermore, it is worth noting that the current development of quantum networks has created an urgent need for on-chip multi-channel QFCs to interconnect diverse quantum systems and achieve multiple manipulations as shown in Fig. 1. To achieve this goal, the pump power per channel must be reduced to milliwatts or even sub-milliwatts. Therefore, developing an electrically pumped ultra-efficient QFC has became an urgent and critical challenge.


In this work, we demonstrate a cavity-enhanced QFC process with periodically poled lithium niobate (PPLN) microring resonator (MRR) on a TFLN chip pumped by a hybrid-integrated laser chip. The achieved ultra-high normalized conversion efficiency of 386,000 \%/W ensures an ulta-low pump power of 360 \(\mu\)W. Thus the QFC could be realized by on-chip pumping using a hybrid integrated distributed feedback (DFB) laser chip, showing the capability of supporting multi-channel QFCs simultaneously. Consequently, a quantum efficiency of 57\% and a noise count of \(\sim\) 7k counts per second (cps) are achieved, demonstrating the feasibility of on-chip QFC between the telecom and visible bands at the single-photon level.


\section{Method and Design}

In this section, we present the method and design of the integrated QFC chip for achieving frequency conversion between the 1550-nm and 630-nm bands using a single-frequency 1064-nm pump laser as shown in Fig. 2(b). A DFB laser chip is edge-coupled to the TFLN chip and the emitting light is efficiently coupled into the waveguide as the pump laser. The directional coupler (DC) efficiently couples the signal and pump light into the MRR. The periodically poled MRR serves as the core component, enabling type-0 quasi-phase matching (QPM) and triple-resonance to facilitate high-efficient nonlinear frequency conversion. Then a SF transimission is designed to optimize the transmission of the sum-frequency (SF) light. Following, we elaborate on the principle of the PPLN MRR nonlinearity, the design of the triple-resonant MRR, the directional coupler and the SF transmission waveguide.

The QFC chip is designed and fabricated on z-cut TFLN wafer, comprising a 600 nm thick MgO-doped lithium niobate layer and a 2 µm thick buried silicon oxide (SiO$_2$) layer on a silicon substrate. We configure the 1550-nm signal, 1064-nm pump, and 631-nm SF lights in quasi-TM fundamental modes to utilize the effective nonlinear coefficient \(d_{33}\). Notably, we employ MgO-doped TFLN to mitigate undesirable dynamic effects which may be caused by photorefractive, thermal, and Kerr nonlinearities in the high quality factor (Q-factor) resonator, thereby significantly enhancing the stability of the MRR.

In this work, our design for the MRR differs significantly from the prior works. Specifically, we select a double-pulley add-drop design over an all-pass design which is applicable to the small-detuning-pump scheme.
In previous work~\cite{28}, telecom-band pump light with 8-nm detuning from the signal is employed to realize SFG in the all-pass configuration. However, this approach introduced significant noise which will destroy the quantum state and make it difficult to achieve QFC at the single-photon level. Generally, to achieve a practical QFC featuring low noise count rate, the three mixing lights usually have significant wavelength differences, such as the SFG process to connect telecom fiber and diamond-color-center-based quantum memories demonstrated here. 

For all-pass MRR~\cite{30,31}, while ensuring that the 1550-nm signal and 1064-nm pump light are efficiently coupled into the microcavity in fundamental modes with optimized coupling strength, achieving efficient output of SF light at 631 nm remains challenging. To address this, we employ a double-pulley add-drop design as shown in Fig. 2(c), which allows independent design for three mixing wavelengths, ensuring efficient coupling for each wavelength and minimizing interference between different modes. The signal and pump light enter the microring resonator through coupler A, and the generated SF light exits through coupler B. This decoupled design introduces additional coupling losses and reduces the Q-factor of the MRR, yet significantly enhances the feasibility of a promising QFC chip.


\begin{figure}
\begin{center}
\begin{tabular}{c}
\includegraphics[height=6.5cm]{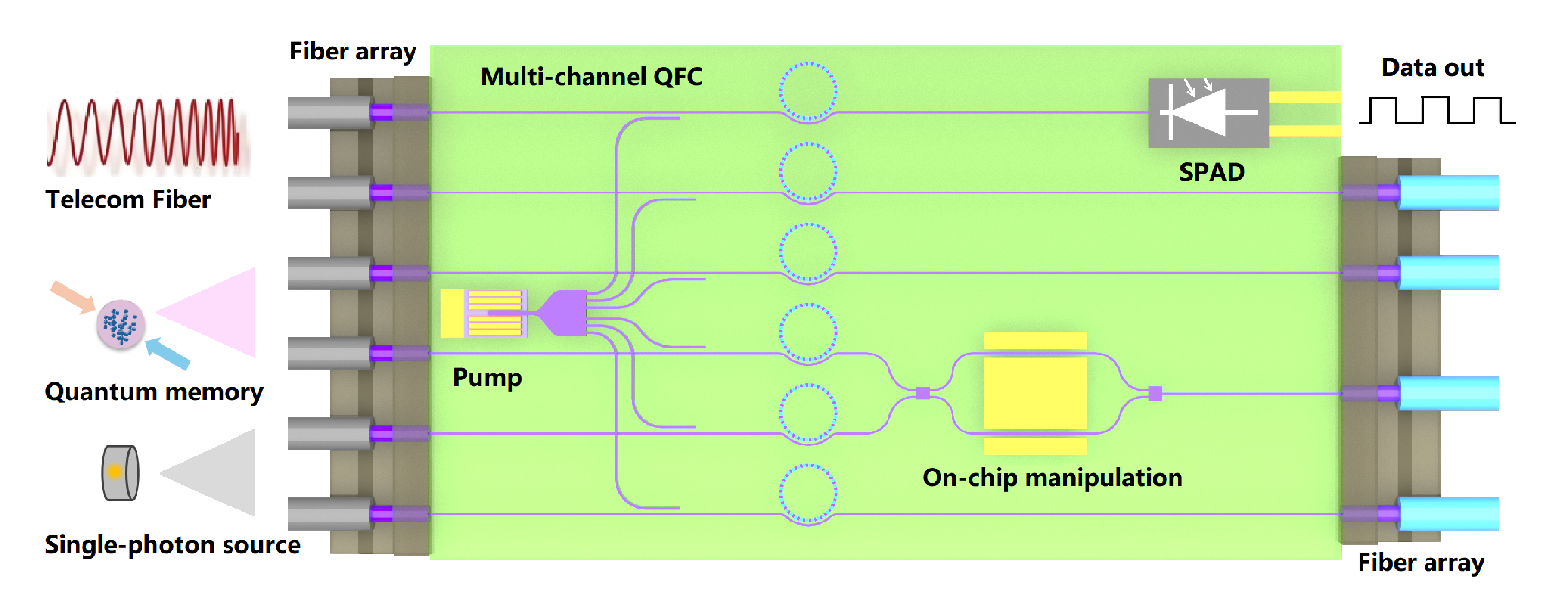}
\end{tabular}
\end{center}
\caption 
{\label{fig:example}
    Schematic diagram of an integrated QFC chip for processing diverse quantum systems, including telecommunication fibers, single-photon sources, quantum memories, single-photon detection and on-chip quantum state manipulation.} 
\end{figure} 

\begin{figure}
\begin{center}
\begin{tabular}{c}
\includegraphics[height=10cm]{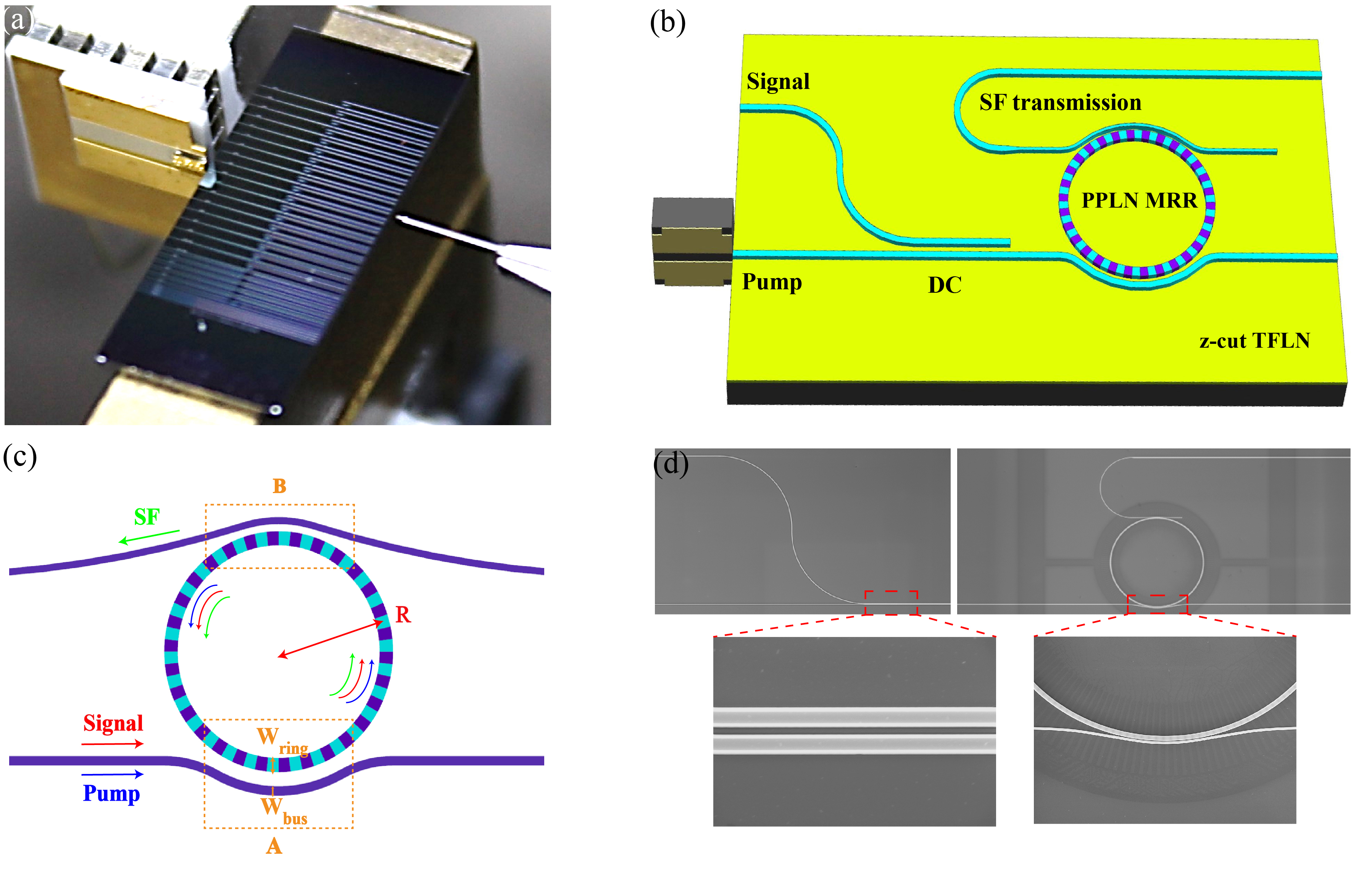}
\end{tabular}
\end{center}
\caption 
{\label{fig:example} Characterization of the electrically pumped QFC on TFLN platform. (a) Photo of the hybrid integration of the DFB laser and TFLN chip. (b) The schematic design of the integrated QFC chip. DC: directional coupler; PPLN MRR: periodically poled lithium niobate microring resonator; SF: sum-frequency. (c) Schematic diagram of the double-pulley add-drop resonator structure. (d) Scanning electron microscopy (SEM) images show an overview of the TFLN chip with the details of the directional coupler and coupler A.} 
\end{figure} 

\subsection{Principle of PPLN MRR}
\label{sect:title}
The Hamiltonian of the QFC process can be expressed as:

\begin{equation}
\widehat{H} = \hbar \omega_{\text{sf}} \widehat{a}_{\text{sf}}^{\dagger} \widehat{a}_{\text{sf}} 
+ \hbar \omega_{\text{s}} \widehat{a}_{\text{s}}^{\dagger} \widehat{a}_{\text{s}} 
+ \hbar \omega_{\text{p}} \widehat{a}_{\text{p}}^{\dagger} \widehat{a}_{\text{p}} 
+ g \left( \widehat{a}_{\text{s}} \widehat{a}_{\text{p}} \widehat{a}_{\text{sf}}^{\dagger} 
+ \widehat{a}_{\text{s}}^{\dagger} \widehat{a}_{\text{p}}^{\dagger} \widehat{a}_{\text{sf}} \right),
\label{eq:1}
\end{equation}
where \(\widehat{a}_{\text{s}}\),  \(\widehat{a}_{\text{p}}\) and  \(\widehat{a}_{\text{sf}}\) represent the annihilation operators of the signal, pump, and SF modes, respectively. $\mathit{g}$ is the nonlinear photon-photon coupling strength, which is given by

\begin{equation}
g \propto \frac{d_{\text{eff}} \zeta}{\sqrt{V_{\text{eff}}}} 
\delta(m_{\text{sf}} - m_{\text{s}} - m_{\text{p}} - M)
\label{eq:2}
\end{equation}
where $d_{\text{eff}}$ represents the effective nonlinear coefficient, $\zeta$ is the mode overlap factor, $V_{\text{eff}}$ denotes the effective mode volume, $m_{j}\ (j = \text{sf}, \text{s}, \text{p})$ corresponds to the azimuthal mode number of the cavity mode, and $M$ signifies the azimuthal polarization period. Quasi-phase matching in a PPLN MRR is achieved when $m_{\text{sf}} - m_{\text{s}} - m_{\text{p}} - M = 0$. According to the theoretical model~\cite{30}, the maximum conversion efficiency of a double-pulley add-drop MRR can be expressed as
\begin{equation}
\eta_{\max} = \left( \frac{r_{\text{sf}}}{1 + r_{\text{sf}}} \right) 
\cdot \left( \frac{r_{\text{s}}}{1 + r_{\text{s}}} \right)
\label{eq:3}
\end{equation}
where $r_\text{sf} = \kappa_{\text{sf, B}}^{2} / (\alpha L+\kappa_{\text{sf, A}}^{2})$
and $r_\text{s} = \kappa_{\text{s, A}}^{2} / (\alpha L+\kappa_{\text{s, B}}^{2}) $
quantify the ratio of coupling strength to internal loss. 
$\kappa_{\text{sf, A}}^{2}$ and $\kappa_{\text{sf, B}}^{2}$ are the coupling strengths of the SF light at couplers A and B, respectively. 
$\kappa_{\text{s, A}}^{2}$ and $\kappa_{\text{s, B}}^{2}$ are
the coupling strengths of the signal light at couplers A and B, respectively. Additionally, $\alpha$ and $L$ denote the propagation loss and the circumference of the microring.

It can be seen from Eq. (3) that the coupling strengths in the couplers and the internal losses within the cavity directly determine the efficiency of frequency conversion. Efficient coupling of the signal into and extraction of SF from the MRR can improve overall conversion efficiency. However, this approach decreases the intrinsic Q factor of the resonator and inevitably increases the required pump power. It is essential to find a balance between the conversion efficiency and pump power. In our design, we firstly ensure that the signal and pump light are efficiently coupled into the fundamental mode of the MRR. Then we tune the coupling strengths of couplers A and B to achieve high-efficiency conversion with a pump power of less than 1 mW.

\subsection{Triple-resonant PPLN MRR}



In this section, we present a detailed introduction of the design for the triple-resonant PPLN MRR. To achieve a high-efficiency QFC with sub-milliwatt pump power, we carefully designed the couplers A and B in the MRR with following fundamental parameters: film thickness of 600 nm, etching depth of 420 nm, microring waveguide width of 1.73 \(\mu\)m, and microring radius of 74 \(\mu\)m. When the triple resonance is achieved, the corresponding azimuthal mode numbers are $m_\text{s}$ = 550, $m_\text{p}$ = 875, and $m_{\text{sf}}$ = 1584. Phase matching is satisfied with $M = m_{\text{sf}} - m_\text{s} - m_\text{p} = 159$. For the phase matching between the bus waveguide and microring, the refractive index matching relation should be satisfied:
\(n_{\text{ring}} \cdot (R + w_{\text{ring}}/4) = n_{\text{wg}} \cdot (R + \text{gap} + w_{\text{ring}}/2 + w_{\text{wg}}/2)\), 
where $n_{\text{ring}}$ and $n_{\text{wg}}$ represent the effective refractive indices of the bus waveguide and microring modes, $w_{\text{wg}}$ and $w_{\text{ring}}$ are the widths of the bus waveguide and microring, $R$ is the microring resonator radius, and \text{gap} is the distance between the microring and the bus waveguide~\cite{31}. 

We employed 3D finite-element simulations to optimize the design of the couplers (See Supplemental Material for details). In coupler A, $w_{\text{wg}}$ and \text{gap} is designed to be 600 nm and 700 nm, respectively. While in coupler B, $w_{\text{wg}}$ and \text{gap} is designed to be 300 nm and 390 nm, respectively. Correspondingly, the coupling strengths for the SF light at couplers A and B are $\kappa_{\text{sf, A}}^{2} \sim 0.005$ and $\kappa_{\text{sf, B}}^{2} \sim 0.05$, while those for the signal light are $\kappa_{\text{s, A}}^{2} \sim 0.03$ and $\kappa_{\text{s, B}}^{2} \sim 0.004$.
Accounting for the typical propagation loss of $\alpha \sim 0.2$ dB/cm and the simulated coupling strength of signal and SF at coupler A and B, the theoretical maximum conversion efficiency is estimated to be $\sim$ 73\%, achievable with a pump power of $\sim$ 100 \(\mu\)W, which is sufficient to satisfy the requirements for multi-channel on-chip QFCs.



\subsection{Directional coupler and SF transmission}

To efficiently couple light at 1550-nm and 1064-nm bands into the PPLN MRR and collect the 631-nm SF light generated, we designed a wavelength division multiplexer and a low-loss SF transmission optical path as shown in Fig. 2(b). The signal and pump beams are combined using a DC consisting of two identical straight waveguides with a width of 0.75 \(\mu\)m and a gap of 0.9 \(\mu\)m.



Furthermore, the output mode of the SF light must be converted into a low-loss fundamental mode and change the transmission direction to facilitate subsequent manipulation. An abrupt taper that rapidly expands the waveguide width from 300 nm to 950 nm is implemented, effectively avoiding mode hybridization while maintaining the TM\textsubscript{0} mode. With an optimized taper length of 4~$\mu$m, the mode crosstalk loss is minimized to 0.24 dB. Then a variable-curvature Euler bend waveguide is used to rotate the transmission direction of SF light without introducing excess loss and mode crosstalk~\cite{32}. The Euler bend is defined by its maximum and minimum radii of curvature, denoted as $R_{\text{max}}$ and $R_{\text{min}}$, respectively. The effective radius  $R_{\text{eff}}$, is defined as the radius of a 90° arc covering the same region. For an SF wavelength of 631 nm and a waveguide width of 950 nm, the optimized parameters are: $R_{\text{max}} = 300~\mu\text{m}$, $R_{\text{min}} = 28.5~\mu\text{m}$, and $R_{\text{eff}} = 50~\mu\text{m}$.
Simulation results show that the Euler bend waveguide achieves a minimal mode loss of 0.01 dB, enabling efficient optical transmission (See Supplemental Material for details).

\section{Results and Discussion}
\label{sect:sections}
To construct the electrically pumped QFC depicted in Fig. 2(a), a single-mode DFB laser chip is integrated with a TFLN photonic circuit using butt-coupling techniques~\cite{ref27,ref26,ref28}. This hybrid integration scheme, illustrated in Fig. 2(b), facilitates efficient optical mode transfer between the components with a coupling efficiency of \(\sim\) 20\%. The DFB laser emits horizontally
polarized pump light at \(\sim\) 1064 nm, exhibiting
a linear temperature-dependent shift of 85.5 pm/°C,
which enables precise wavelength tuning through thermal adjustment. This is critical as the MRR exhibits narrow resonance peaks with linewidths of GHz. In view of the polarization-mode mismatch, the laser chip is rotated by 90° to ensure efficient coupling with the TFLN chip. The periodic poling of the MRR is achieved with an external electric field poling method~\cite{29}. Subsequently, the waveguide structures are fabricated using electron beam lithography (EBL) and inductively coupled plasma (ICP) etching, with the scanning electron microscopy (SEM) images of the structures shown in Fig. 2(d).

\subsection{Characterization of the PPLN MRR and DC}
To measure the Q-factors of the three frequencies, a dedicated design is developed on the same chip without SF transmission path. The performances of the fabricated PPLN MRR are characterized by separately injecting tunable single-frequency laser at the pump and signal band through coupler A, and tunable laser at SF band through coupler B. For coupler B on the test chip, the straight waveguides at the input and output ports are of the same width with the bus waveguide of 300 nm. 

The resonance spectra of the signal, pump, and SF light are presented in Fig. 3(a). All three wavelengths operate in the over-coupled regime, with the intrinsic Q factors of the signal, pump, and SF calculated as \(Q_{\text{s},0} = 1.01 \times 10^{6}\), \(Q_{\text{p},0} = 3.29 \times 10^{6}\), \(Q_{\text{sf},0} = 8.93 \times 10^{5}\), respectively, and the loaded Q factors as \(Q_{\text{s},\text{l}} = 1.46 \times 10^{5}\), \(Q_{\text{p},\text{l}} = 5.26 \times 10^{5}\), \(Q_{\text{sf},\text{l}} = 1.64 \times 10^{5}\), respectively. The intrinsic Q factors of the signal and pump lights are determined by the intracavity propagation loss and the coupling loss in coupler B, while the intrinsic Q factor of the SF light is determined by the intracavity propagation loss and the coupling loss in coupler A. Using the expression 
\(
\eta_{\max} = (1 - \frac{Q_{\text{s},\text{l}}}{\ Q_{\text{s},\text{0}}}) (1 - \frac{Q_{\text{sf},\text{l}}}{Q_{\text{sf},\text{0}}})
\) derived from Eq. (3),
the maximum conversion efficiency of the PPLN MRR is calculated to be \(\sim\) 70\%, demonstrating excellent agreement with the theoretical design.


Additionally, an independent DC structures with identical parameters with the design is fabricated to characterize the performance. Figure 3(b) shows the simulated and measured wavelength-dependent coupling efficiency of the signal, defined as the fraction of optical power transferred from the input waveguide to the adjacent waveguide. With a DC length of 450 \(\mu\)m, the coupling efficiency reaches \(\sim\) 98\% at 1533 nm which is the QPM wavelength of our integrated QFC chip. The coupling loss of the pump light is characterized similarly and measured to be \(\sim\) 1.2\%. The signal and pump beams are effectively combined with small insertion losses.

\begin{figure}
\begin{center}
\begin{tabular}{c}
\hspace*{-0.8cm}
\includegraphics[height=6.2cm]{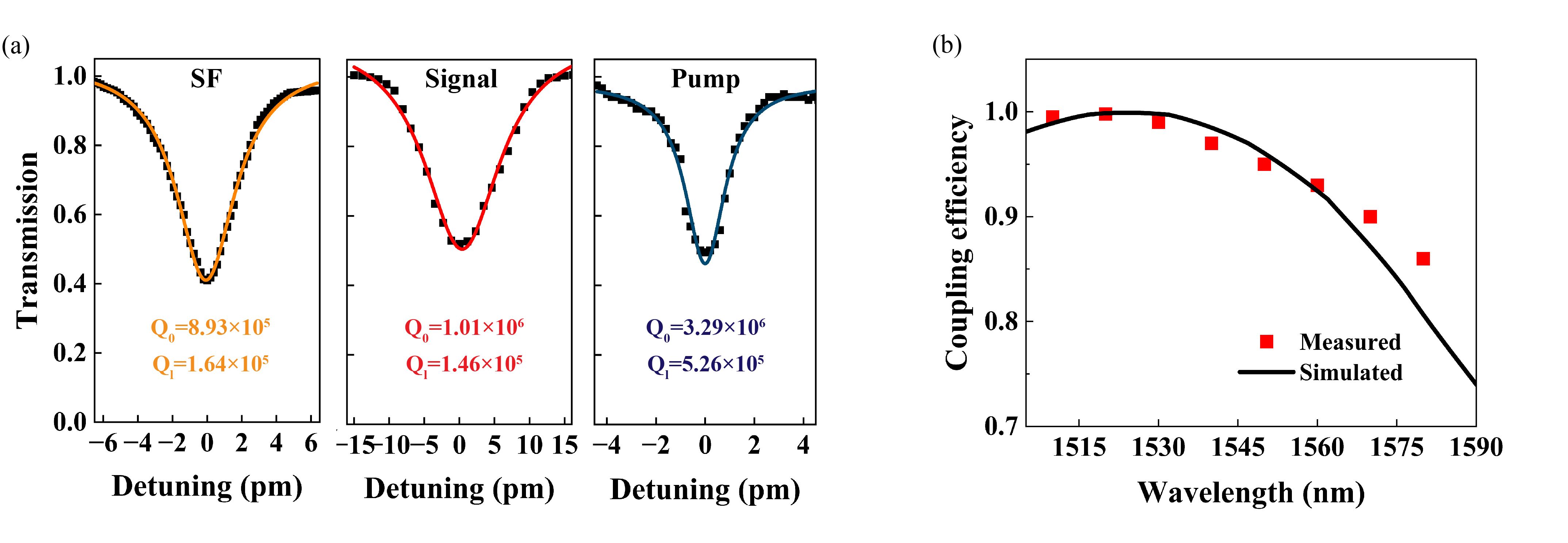}
\end{tabular}
\end{center}
\caption 
{ \label{fig:example}
(a) Resonance spectra of the signal, pump, and SF. (b) Measured and simulated coupling efficiency of DC at different signal wavelengths.} 
\end{figure} 

\subsection{Characterization of the integrated QFC chip}
The experimental setup for characterizing the integrated QFC chip is shown in Fig. 4(a). A variable optical attenuator (VOA) is used to adjust the power of the signal. A polarization controller (PC) is employed to align its polarization before coupling into the waveguide via a lensed fiber with a mode field diameter of $2.5 \pm 0.5\ \mu$m. The dual-end coupling measurement shows that the coupling efficiency of the signal between the lensed fiber and the TFLN chip is \(\sim\) 30\%. The generated SF light at $\sim$ 628 nm is collected by an aspheric lens (AL), passing through a spatial filtering system (Filters), then detected by a power meter. The wavelength of the DFB laser around 1064 nm is finely tuned using a high-precision temperature control module to align with the cavity resonance, ensuring strong coupling and stable buildup of the pump light inside the cavity. The signal wavelength is then scanned around 1533 nm, and the peak intensity of the collected SF light is recorded. 
The temperature of the TFLN chip is set at 40.5 $^{\circ}$C and fine-tuned with a precision of $\sim 0.02$ K to compensate for slight resonance shifts induced by thermo-optic and photo-refractive effects while varying pump power.
Finally, the signal light is turned off, and the output photons are filtered and coupled into a multimode fiber before being detected by a silicon single-photon avalanche diode (SPAD). The on-chip noise photon count is derived after accounting for the SPAD detection efficiency, background noise, and transmission losses in the free-space optics.

\begin{figure}
\begin{center}
\begin{tabular}{c}
\hspace*{-0.6cm}
\includegraphics[height=12.5cm]{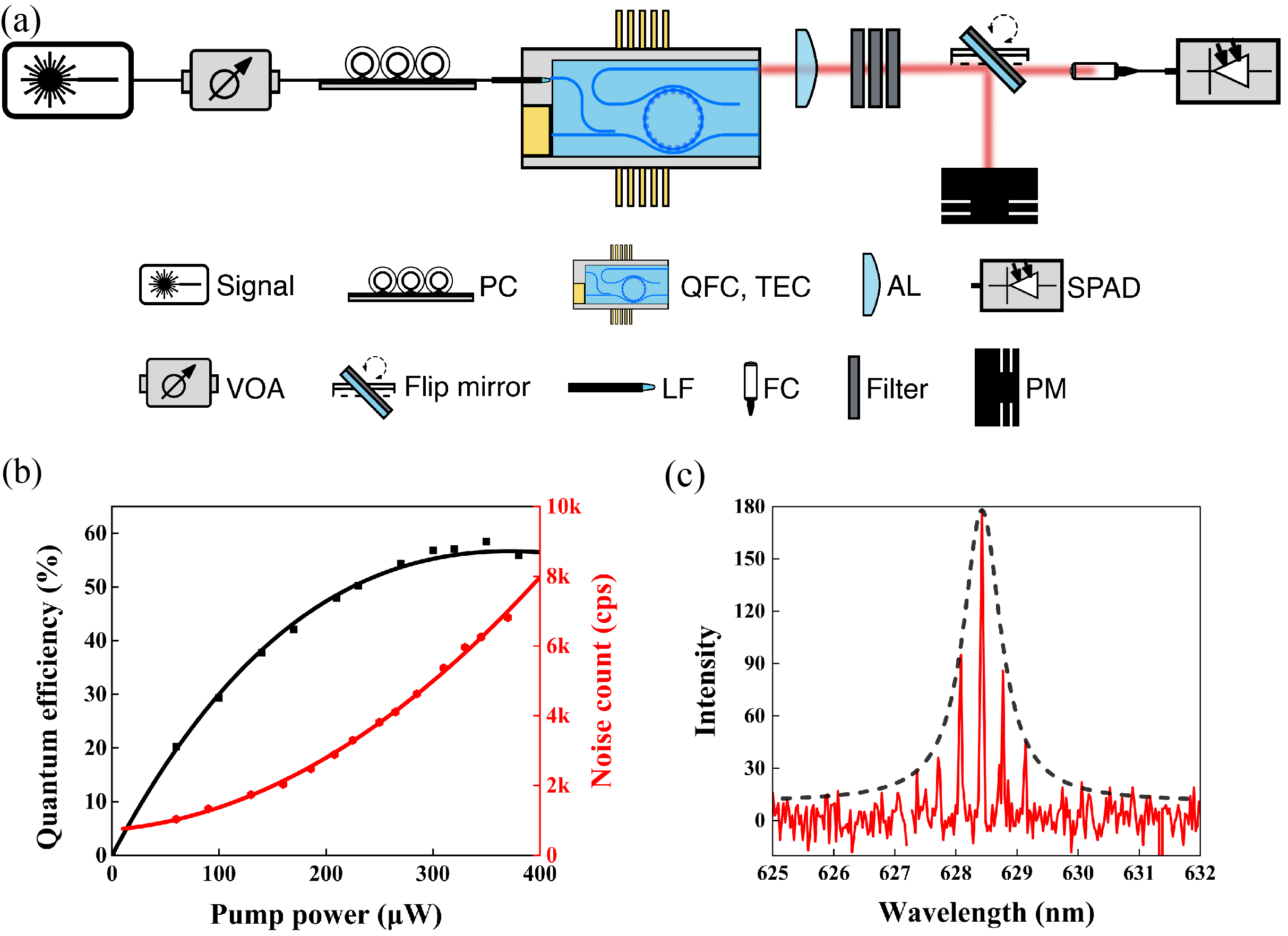}
\end{tabular}
\end{center}
\caption 
{\label{fig:example}
(a) Experimental setup for measuring the quantum efficiency and noise of the QFC chip. VOA, variable optical attenuator; PC, polarization controller; TEC, thermoelectric cooler; AL, aspherical lens; Filter, including 600-nm longpass filter, 800-nm shortpass filter, and 4-nm bandpass filter; PM, power meter; FC, fiber coupler; SPAD, silicon single-photon avalanche photodiode; LF, lensed fiber. (b) Quantum conversion efficiency (blue) and on-chip noise (red) as a function of the pump power. (c) Noise spectrum of the QFC chip. The dashed line represents the envelope of the nonlinear noise.} 
\end{figure} 

Figure 4(b) shows the quantum efficiency (QE) and on-chip noise of our QFC chip as a function of pump power. The QE is calculated as \(\eta = P_{\text{sf}}\lambda_{\text{sf}}/P_{\text{s}}\lambda_{\text{s}}\), where \(P_{\text{s}}\) is the coupled signal power on-chip (\(\sim 20\ \text{nW}\)) and \(P_{\text{sf}}\) is the output power of the generated SF. Based on the fitting curve, the maximum QE occurs at a pump power of \(360\ \mu\text{W}\), corresponding to a normalized conversion efficiency of 386,000 \%/W. Consequently, an on-chip maximum QE of 57\% is achieved with a noise count of \(\sim\) 7k cps. The discrepancy between the measured on-chip QE and the theoretical QE of the PPLN MRR are attributed to several factors, including the signal loss introduced by the DC, the coupling of the signal light into higher-order cavity modes, and the transmission and collection losses of the SF light. The pump powers corresponding to the maximum QE, as obtained from both simulation and experimental results, are in close agreement. Theoretically, pump power scales inversely with the product of the intrinsic Q-factors associated with signal, pump and SF~\cite{30}. Consequently, the observed discrepancies likely stem from variations between the design and fabrication parameters, including the propagation losses and the coupling strengths at coupler A and B.





The current on-chip QE is comparable to the traditional straight waveguide schemes~\cite{van2022entangling,liu2024creation,PhysRevApplied.14.034035}, and the noise count has been reduced by two orders of magnitude compared with the previous work based on PPLN MRR~\cite{28}, ensuring the realization of high-performance single-photon QFC. Additionally, considering the output power of the DFB chip and the coupling efficiency between the DFB and the TFLN chip, our design could support QFCs for more than 10 channels. By optimizing the coupling design at the input port of the TFLN chip, future implementations could achieve dozens of channels. Furthermore, the flexibility of domain engineering allows tailored phase matching, facilitating seamless interconnections and quantum state manipulations across diverse quantum systems operating at different wavelengths.


Another important characteristic of a QFC device is noise photon generation, which is primarily caused by the strong pump through spontaneous Raman scattering (SRS) and spontaneous parametric down-conversion (SPDC) process. Furthermore, second harmonic generation (SHG) of the pump laser induced SPDC and SRS processes may also contribute non-negligible amount of noise photons.
To further analyze the generation mechanism, the noise photons are coupled into a multimode fiber and then sent to a spectrometer. Distinct noise peaks are observed in the spectrum as shown in in Fig. 4(c), aligning with resonance peaks of the PPLN MRR. While SRS and SPDC processes typically exhibit a flat spectral profile across the SFG bandwidth, resulting in uniformly intense peaks at the cavity output, significant variations in noise peak intensities revealed in Fig. 4(c). This indicates that the noise predominantly originates from photons generated in the signal band, which are subsequently converted to the SF band via SFG. Moreover, it is worth noting that the PPLN MRR could significantly suppresses the on-chip nonlinear noise, which is a unique feature compared with traditional QFC devices.


\section{Conclusion}

By integrating DFB laser with TFLN chip, we have successfully developed an electrically pumped, integrated and scalable QFC chip in a compact hybrid platform (12 \(\times\) 12 $\rm{mm}^2$). As a crucial element in developing the hybrid integration, the realization of PPLN MRR with ultra-high normalized conversion efficiency of 386,000 \%/W guarantees an ultra-low pump power of 360 \(\mu\)W per channel. By injecting current into the chip, the telecom band single photon is converted to the visible band matching the diamond-color-center-based quantum memories. An on-chip quantum efficiency of 57\% and a noise count of \(\sim\) 7k cps are achieved. Such an integrated and scalable QFC chip shows the capability of supporting multi-channel QFCs simultaneously, which would significantly advancing the integration of quantum network and the development of chip-scale quantum optical systems as shown in Fig. (1). 

To meet the future demands of quantum science and technology, on-chip multi-channel QFCs become an essential requirement, particularly for multimode and multinode quantum repeaters, and distributed quantum computing. Significantly, when integrated with on-chip quantum state manipulation and single-photon detection, fully on-chip integrated quantum chips will lead to a quantum leap in quantum technologies, profoundly advancing quantum communication, quantum computing, and quantum metrology.

\subsection* {Acknowledgments}
We thank Zhi-Chuan Niu, Yu Zhang and Cheng-Ao Yang for useful discussions and the test of DFB-TFLN hybrid integration; Juanjuan Lu for insightful discussions on the theory of MRR nonlinearity. This work is supported by the Key R\(\And\)D Plan of Shandong Province (Grant No. 2024CXPT083, No. 2021ZDPT01), the Innovation Program for Quantum Science and Technology (Grant No. 2021ZD0300800, No. 2021ZD0300802), the National Natural Science Foundation of China (Grant No. T2125010), Natural Science Foundation of Shandong Province (Grant No. ZR2021LLZ013, ZR2022LLZ009, ZR2022LLZ010). Q. Z. were supported by the New Cornerstone Science Foundation through the Xplorer Prize, the SAICT Experts Program, the Taishan Scholar Program of Shandong Province, and Quancheng Industrial Experts Program. M.-Y. Z. were supported by the Taishan Scholar Program of Shandong Province and Haiyou Plan Project of Jinan.
\bibliography{report}

\begin{thebibliography}{44}%
\makeatletter
\providecommand \@ifxundefined [1]{%
 \@ifx{#1\undefined}
}%
\providecommand \@ifnum [1]{%
 \ifnum #1\expandafter \@firstoftwo
 \else \expandafter \@secondoftwo
 \fi
}%
\providecommand \@ifx [1]{%
 \ifx #1\expandafter \@firstoftwo
 \else \expandafter \@secondoftwo
 \fi
}%
\providecommand \natexlab [1]{#1}%
\providecommand \enquote  [1]{``#1''}%
\providecommand \bibnamefont  [1]{#1}%
\providecommand \bibfnamefont [1]{#1}%
\providecommand \citenamefont [1]{#1}%
\providecommand \href@noop [0]{\@secondoftwo}%
\providecommand \href [0]{\begingroup \@sanitize@url \@href}%
\providecommand \@href[1]{\@@startlink{#1}\@@href}%
\providecommand \@@href[1]{\endgroup#1\@@endlink}%
\providecommand \@sanitize@url [0]{\catcode `\\12\catcode `\$12\catcode `\&12\catcode `\#12\catcode `\^12\catcode `\_12\catcode `\%12\relax}%
\providecommand \@@startlink[1]{}%
\providecommand \@@endlink[0]{}%
\providecommand \url  [0]{\begingroup\@sanitize@url \@url }%
\providecommand \@url [1]{\endgroup\@href {#1}{\urlprefix }}%
\providecommand \urlprefix  [0]{URL }%
\providecommand \Eprint [0]{\href }%
\providecommand \doibase [0]{https://doi.org/}%
\providecommand \selectlanguage [0]{\@gobble}%
\providecommand \bibinfo  [0]{\@secondoftwo}%
\providecommand \bibfield  [0]{\@secondoftwo}%
\providecommand \translation [1]{[#1]}%
\providecommand \BibitemOpen [0]{}%
\providecommand \bibitemStop [0]{}%
\providecommand \bibitemNoStop [0]{.\EOS\space}%
\providecommand \EOS [0]{\spacefactor3000\relax}%
\providecommand \BibitemShut  [1]{\csname bibitem#1\endcsname}%
\let\auto@bib@innerbib\@empty
\bibitem [{\citenamefont {Kumar}(1990)}]{1}%
  \BibitemOpen
  \bibfield  {author} {\bibinfo {author} {\bibfnamefont {P.}~\bibnamefont {Kumar}},\ }\bibfield  {title} {\bibinfo {title} {Quantum frequency conversion},\ }\href@noop {} {\bibfield  {journal} {\bibinfo  {journal} {Opt. Lett.}\ }\textbf {\bibinfo {volume} {15}},\ \bibinfo {pages} {1476} (\bibinfo {year} {1990})}\BibitemShut {NoStop}%
\bibitem [{\citenamefont {Huang}\ and\ \citenamefont {Kumar}(1992)}]{2}%
  \BibitemOpen
  \bibfield  {author} {\bibinfo {author} {\bibfnamefont {J.~M.}\ \bibnamefont {Huang}}\ and\ \bibinfo {author} {\bibfnamefont {P.}~\bibnamefont {Kumar}},\ }\bibfield  {title} {\bibinfo {title} {Observation of quantum frequency conversion},\ }\href@noop {} {\bibfield  {journal} {\bibinfo  {journal} {Phys. Rev. Lett.}\ }\textbf {\bibinfo {volume} {68}},\ \bibinfo {pages} {2153} (\bibinfo {year} {1992})}\BibitemShut {NoStop}%
\bibitem [{\citenamefont {Ikuta}\ \emph {et~al.}(2011)\citenamefont {Ikuta}, \citenamefont {Kusaka}, \citenamefont {Kitano} \emph {et~al.}}]{15}%
  \BibitemOpen
  \bibfield  {author} {\bibinfo {author} {\bibfnamefont {R.}~\bibnamefont {Ikuta}}, \bibinfo {author} {\bibfnamefont {Y.}~\bibnamefont {Kusaka}}, \bibinfo {author} {\bibfnamefont {T.}~\bibnamefont {Kitano}}, \emph {et~al.},\ }\bibfield  {title} {\bibinfo {title} {Wide-band quantum interface for visible-to telecommunication wavelength conversion},\ }\href {https://doi.org/10.1038/ncomms1544} {\bibfield  {journal} {\bibinfo  {journal} {Nat. Commun.}\ }\textbf {\bibinfo {volume} {2}},\ \bibinfo {pages} {537} (\bibinfo {year} {2011})}\BibitemShut {NoStop}%
\bibitem [{\citenamefont {Albrecht}\ \emph {et~al.}(2014)\citenamefont {Albrecht}, \citenamefont {Farrera}, \citenamefont {Fernandez-Gonzalvo} \emph {et~al.}}]{4}%
  \BibitemOpen
  \bibfield  {author} {\bibinfo {author} {\bibfnamefont {B.}~\bibnamefont {Albrecht}}, \bibinfo {author} {\bibfnamefont {P.}~\bibnamefont {Farrera}}, \bibinfo {author} {\bibfnamefont {X.}~\bibnamefont {Fernandez-Gonzalvo}}, \emph {et~al.},\ }\bibfield  {title} {\bibinfo {title} {A waveguide frequency converter connecting rubidium-based quantum memories to the telecom c-band},\ }\href {https://doi.org/10.1038/ncomms4376} {\bibfield  {journal} {\bibinfo  {journal} {Nat. Commun.}\ }\textbf {\bibinfo {volume} {5}},\ \bibinfo {pages} {3376} (\bibinfo {year} {2014})}\BibitemShut {NoStop}%
\bibitem [{\citenamefont {Maring}\ \emph {et~al.}(2017)\citenamefont {Maring}, \citenamefont {Farrera}, \citenamefont {Kutluer} \emph {et~al.}}]{3}%
  \BibitemOpen
  \bibfield  {author} {\bibinfo {author} {\bibfnamefont {N.}~\bibnamefont {Maring}}, \bibinfo {author} {\bibfnamefont {P.}~\bibnamefont {Farrera}}, \bibinfo {author} {\bibfnamefont {K.}~\bibnamefont {Kutluer}}, \emph {et~al.},\ }\bibfield  {title} {\bibinfo {title} {Photonic quantum state transfer between a cold atomic gas and a crystal},\ }\href {https://doi.org/10.1038/nature24468} {\bibfield  {journal} {\bibinfo  {journal} {Nature}\ }\textbf {\bibinfo {volume} {551}},\ \bibinfo {pages} {485} (\bibinfo {year} {2017})}\BibitemShut {NoStop}%
\bibitem [{\citenamefont {Walker}\ \emph {et~al.}(2018)\citenamefont {Walker}, \citenamefont {Miyanishi}, \citenamefont {Ikuta} \emph {et~al.}}]{8}%
  \BibitemOpen
  \bibfield  {author} {\bibinfo {author} {\bibfnamefont {T.}~\bibnamefont {Walker}}, \bibinfo {author} {\bibfnamefont {K.}~\bibnamefont {Miyanishi}}, \bibinfo {author} {\bibfnamefont {R.}~\bibnamefont {Ikuta}}, \emph {et~al.},\ }\bibfield  {title} {\bibinfo {title} {Long-distance single photon transmission from a trapped ion via quantum frequency conversion},\ }\href {https://doi.org/10.1103/PhysRevLett.120.203601} {\bibfield  {journal} {\bibinfo  {journal} {Phys. Rev. Lett.}\ }\textbf {\bibinfo {volume} {120}},\ \bibinfo {pages} {203601} (\bibinfo {year} {2018})}\BibitemShut {NoStop}%
\bibitem [{\citenamefont {Bock}\ \emph {et~al.}(2018)\citenamefont {Bock}, \citenamefont {Eich}, \citenamefont {Kucera} \emph {et~al.}}]{14}%
  \BibitemOpen
  \bibfield  {author} {\bibinfo {author} {\bibfnamefont {M.}~\bibnamefont {Bock}}, \bibinfo {author} {\bibfnamefont {P.}~\bibnamefont {Eich}}, \bibinfo {author} {\bibfnamefont {S.}~\bibnamefont {Kucera}}, \emph {et~al.},\ }\bibfield  {title} {\bibinfo {title} {High-fidelity entanglement between a trapped ion and a telecom photon via quantum frequency conversion},\ }\href {https://doi.org/10.1038/s41467-018-04339-w} {\bibfield  {journal} {\bibinfo  {journal} {Nat. Commun.}\ }\textbf {\bibinfo {volume} {9}},\ \bibinfo {pages} {1998} (\bibinfo {year} {2018})}\BibitemShut {NoStop}%
\bibitem [{\citenamefont {Maring}\ \emph {et~al.}(2018)\citenamefont {Maring}, \citenamefont {Lago-Rivera}, \citenamefont {Lenhard} \emph {et~al.}}]{5}%
  \BibitemOpen
  \bibfield  {author} {\bibinfo {author} {\bibfnamefont {N.}~\bibnamefont {Maring}}, \bibinfo {author} {\bibfnamefont {D.}~\bibnamefont {Lago-Rivera}}, \bibinfo {author} {\bibfnamefont {A.}~\bibnamefont {Lenhard}}, \emph {et~al.},\ }\bibfield  {title} {\bibinfo {title} {Quantum frequency conversion of memory-compatible single photons from 606 nm to the telecom {C}-band},\ }\href {https://doi.org/10.1364/OPTICA.5.000507} {\bibfield  {journal} {\bibinfo  {journal} {Optica}\ }\textbf {\bibinfo {volume} {5}},\ \bibinfo {pages} {507} (\bibinfo {year} {2018})}\BibitemShut {NoStop}%
\bibitem [{\citenamefont {van Leent}\ \emph {et~al.}(2020)\citenamefont {van Leent}, \citenamefont {Bock}, \citenamefont {Garthoff}, \citenamefont {Redeker}, \citenamefont {Zhang}, \citenamefont {Bauer}, \citenamefont {Rosenfeld}, \citenamefont {Becher},\ and\ \citenamefont {Weinfurter}}]{van2020long}%
  \BibitemOpen
  \bibfield  {author} {\bibinfo {author} {\bibfnamefont {T.}~\bibnamefont {van Leent}}, \bibinfo {author} {\bibfnamefont {M.}~\bibnamefont {Bock}}, \bibinfo {author} {\bibfnamefont {R.}~\bibnamefont {Garthoff}}, \bibinfo {author} {\bibfnamefont {K.}~\bibnamefont {Redeker}}, \bibinfo {author} {\bibfnamefont {W.}~\bibnamefont {Zhang}}, \bibinfo {author} {\bibfnamefont {T.}~\bibnamefont {Bauer}}, \bibinfo {author} {\bibfnamefont {W.}~\bibnamefont {Rosenfeld}}, \bibinfo {author} {\bibfnamefont {C.}~\bibnamefont {Becher}},\ and\ \bibinfo {author} {\bibfnamefont {H.}~\bibnamefont {Weinfurter}},\ }\bibfield  {title} {\bibinfo {title} {Long-distance distribution of atom-photon entanglement at telecom wavelength},\ }\href@noop {} {\bibfield  {journal} {\bibinfo  {journal} {Phys. Rev. Lett.}\ }\textbf {\bibinfo {volume} {124}},\ \bibinfo {pages} {010510} (\bibinfo {year} {2020})}\BibitemShut {NoStop}%
\bibitem [{\citenamefont {Yu}\ \emph {et~al.}(2020)\citenamefont {Yu}, \citenamefont {Ma}, \citenamefont {Luo} \emph {et~al.}}]{10}%
  \BibitemOpen
  \bibfield  {author} {\bibinfo {author} {\bibfnamefont {Y.}~\bibnamefont {Yu}}, \bibinfo {author} {\bibfnamefont {F.}~\bibnamefont {Ma}}, \bibinfo {author} {\bibfnamefont {X.~Y.}\ \bibnamefont {Luo}}, \emph {et~al.},\ }\bibfield  {title} {\bibinfo {title} {Entanglement of two quantum memories via fibers over dozens of kilometers},\ }\href {https://doi.org/10.1038/s41586-020-1976-7} {\bibfield  {journal} {\bibinfo  {journal} {Nature}\ }\textbf {\bibinfo {volume} {578}},\ \bibinfo {pages} {240} (\bibinfo {year} {2020})}\BibitemShut {NoStop}%
\bibitem [{\citenamefont {Zaske}\ \emph {et~al.}(2012)\citenamefont {Zaske}, \citenamefont {Lenhard}, \citenamefont {Kessler} \emph {et~al.}}]{6}%
  \BibitemOpen
  \bibfield  {author} {\bibinfo {author} {\bibfnamefont {S.}~\bibnamefont {Zaske}}, \bibinfo {author} {\bibfnamefont {A.}~\bibnamefont {Lenhard}}, \bibinfo {author} {\bibfnamefont {C.~A.}\ \bibnamefont {Kessler}}, \emph {et~al.},\ }\bibfield  {title} {\bibinfo {title} {Visible-to-telecom quantum frequency conversion of light from a single quantum emitter},\ }\href {https://doi.org/10.1103/PhysRevLett.109.147404} {\bibfield  {journal} {\bibinfo  {journal} {Phys. Rev. Lett.}\ }\textbf {\bibinfo {volume} {109}},\ \bibinfo {pages} {147404} (\bibinfo {year} {2012})}\BibitemShut {NoStop}%
\bibitem [{\citenamefont {Ates}\ \emph {et~al.}(2012)\citenamefont {Ates}, \citenamefont {Agha}, \citenamefont {Gulinatti} \emph {et~al.}}]{7}%
  \BibitemOpen
  \bibfield  {author} {\bibinfo {author} {\bibfnamefont {S.}~\bibnamefont {Ates}}, \bibinfo {author} {\bibfnamefont {I.}~\bibnamefont {Agha}}, \bibinfo {author} {\bibfnamefont {A.}~\bibnamefont {Gulinatti}}, \emph {et~al.},\ }\bibfield  {title} {\bibinfo {title} {Two-photon interference using background-free quantum frequency conversion of single photons emitted by an {InAs} quantum dot},\ }\href {https://doi.org/10.1103/PhysRevLett.109.147405} {\bibfield  {journal} {\bibinfo  {journal} {Phys. Rev. Lett.}\ }\textbf {\bibinfo {volume} {109}},\ \bibinfo {pages} {147405} (\bibinfo {year} {2012})}\BibitemShut {NoStop}%
\bibitem [{\citenamefont {De~Greve}\ \emph {et~al.}(2012)\citenamefont {De~Greve}, \citenamefont {Yu}, \citenamefont {McMahon}, \citenamefont {Pelc}, \citenamefont {Natarajan}, \citenamefont {Kim}, \citenamefont {Abe}, \citenamefont {Maier}, \citenamefont {Schneider}, \citenamefont {Kamp} \emph {et~al.}}]{de2012quantum}%
  \BibitemOpen
  \bibfield  {author} {\bibinfo {author} {\bibfnamefont {K.}~\bibnamefont {De~Greve}}, \bibinfo {author} {\bibfnamefont {L.}~\bibnamefont {Yu}}, \bibinfo {author} {\bibfnamefont {P.~L.}\ \bibnamefont {McMahon}}, \bibinfo {author} {\bibfnamefont {J.~S.}\ \bibnamefont {Pelc}}, \bibinfo {author} {\bibfnamefont {C.~M.}\ \bibnamefont {Natarajan}}, \bibinfo {author} {\bibfnamefont {N.~Y.}\ \bibnamefont {Kim}}, \bibinfo {author} {\bibfnamefont {E.}~\bibnamefont {Abe}}, \bibinfo {author} {\bibfnamefont {S.}~\bibnamefont {Maier}}, \bibinfo {author} {\bibfnamefont {C.}~\bibnamefont {Schneider}}, \bibinfo {author} {\bibfnamefont {M.}~\bibnamefont {Kamp}}, \emph {et~al.},\ }\bibfield  {title} {\bibinfo {title} {Quantum-dot spin--photon entanglement via frequency downconversion to telecom wavelength},\ }\href@noop {} {\bibfield  {journal} {\bibinfo  {journal} {Nature}\ }\textbf {\bibinfo {volume} {491}},\ \bibinfo {pages} {421} (\bibinfo {year} {2012})}\BibitemShut {NoStop}%
\bibitem [{\citenamefont {Singh}\ \emph {et~al.}(2019)\citenamefont {Singh}, \citenamefont {Li}, \citenamefont {Liu}, \citenamefont {Yu}, \citenamefont {Lu}, \citenamefont {Schneider}, \citenamefont {H{\"o}fling}, \citenamefont {Lawall}, \citenamefont {Verma}, \citenamefont {Mirin} \emph {et~al.}}]{singh2019quantum}%
  \BibitemOpen
  \bibfield  {author} {\bibinfo {author} {\bibfnamefont {A.}~\bibnamefont {Singh}}, \bibinfo {author} {\bibfnamefont {Q.}~\bibnamefont {Li}}, \bibinfo {author} {\bibfnamefont {S.}~\bibnamefont {Liu}}, \bibinfo {author} {\bibfnamefont {Y.}~\bibnamefont {Yu}}, \bibinfo {author} {\bibfnamefont {X.}~\bibnamefont {Lu}}, \bibinfo {author} {\bibfnamefont {C.}~\bibnamefont {Schneider}}, \bibinfo {author} {\bibfnamefont {S.}~\bibnamefont {H{\"o}fling}}, \bibinfo {author} {\bibfnamefont {J.}~\bibnamefont {Lawall}}, \bibinfo {author} {\bibfnamefont {V.}~\bibnamefont {Verma}}, \bibinfo {author} {\bibfnamefont {R.}~\bibnamefont {Mirin}}, \emph {et~al.},\ }\bibfield  {title} {\bibinfo {title} {Quantum frequency conversion of a quantum dot single-photon source on a nanophotonic chip},\ }\href@noop {} {\bibfield  {journal} {\bibinfo  {journal} {Optica}\ }\textbf {\bibinfo {volume} {6}},\ \bibinfo {pages} {563} (\bibinfo {year} {2019})}\BibitemShut {NoStop}%
\bibitem [{\citenamefont {You}\ \emph {et~al.}(2022)\citenamefont {You}, \citenamefont {Zheng}, \citenamefont {Chen}, \citenamefont {Liu}, \citenamefont {Qin}, \citenamefont {Xu}, \citenamefont {Ge}, \citenamefont {Chung}, \citenamefont {Qiao}, \citenamefont {Jiang}, \citenamefont {Zhong}, \citenamefont {Chen}, \citenamefont {Wang}, \citenamefont {He}, \citenamefont {Xie}, \citenamefont {Li}, \citenamefont {You}, \citenamefont {Schneider}, \citenamefont {Yin}, \citenamefont {Chen}, \citenamefont {Benyoucef}, \citenamefont {Huo}, \citenamefont {Höfling}, \citenamefont {Zhang}, \citenamefont {Lu},\ and\ \citenamefont {Pan}}]{Li:18}%
  \BibitemOpen
  \bibfield  {author} {\bibinfo {author} {\bibfnamefont {X.}~\bibnamefont {You}}, \bibinfo {author} {\bibfnamefont {M.~Y.}\ \bibnamefont {Zheng}}, \bibinfo {author} {\bibfnamefont {S.}~\bibnamefont {Chen}}, \bibinfo {author} {\bibfnamefont {R.-Z.}\ \bibnamefont {Liu}}, \bibinfo {author} {\bibfnamefont {J.}~\bibnamefont {Qin}}, \bibinfo {author} {\bibfnamefont {M.-C.}\ \bibnamefont {Xu}}, \bibinfo {author} {\bibfnamefont {Z.-X.}\ \bibnamefont {Ge}}, \bibinfo {author} {\bibfnamefont {T.-H.}\ \bibnamefont {Chung}}, \bibinfo {author} {\bibfnamefont {Y.-K.}\ \bibnamefont {Qiao}}, \bibinfo {author} {\bibfnamefont {Y.-F.}\ \bibnamefont {Jiang}}, \bibinfo {author} {\bibfnamefont {H.-S.}\ \bibnamefont {Zhong}}, \bibinfo {author} {\bibfnamefont {M.-C.}\ \bibnamefont {Chen}}, \bibinfo {author} {\bibfnamefont {H.}~\bibnamefont {Wang}}, \bibinfo {author} {\bibfnamefont {Y.-M.}\ \bibnamefont {He}}, \bibinfo {author} {\bibfnamefont {X.-P.}\ \bibnamefont {Xie}}, \bibinfo {author} {\bibfnamefont {H.}~\bibnamefont {Li}},
  \bibinfo {author} {\bibfnamefont {L.-X.}\ \bibnamefont {You}}, \bibinfo {author} {\bibfnamefont {C.}~\bibnamefont {Schneider}}, \bibinfo {author} {\bibfnamefont {J.}~\bibnamefont {Yin}}, \bibinfo {author} {\bibfnamefont {T.-Y.}\ \bibnamefont {Chen}}, \bibinfo {author} {\bibfnamefont {M.}~\bibnamefont {Benyoucef}}, \bibinfo {author} {\bibfnamefont {Y.-H.}\ \bibnamefont {Huo}}, \bibinfo {author} {\bibfnamefont {S.}~\bibnamefont {Höfling}}, \bibinfo {author} {\bibfnamefont {Q.}~\bibnamefont {Zhang}}, \bibinfo {author} {\bibfnamefont {C.-Y.}\ \bibnamefont {Lu}},\ and\ \bibinfo {author} {\bibfnamefont {J.-W.}\ \bibnamefont {Pan}},\ }\bibfield  {title} {\bibinfo {title} {Quantum interference with independent single-photon sources over 300 km fiber},\ }\href@noop {} {\bibfield  {journal} {\bibinfo  {journal} {Adv. Photon.}\ }\textbf {\bibinfo {volume} {4}},\ \bibinfo {pages} {066003} (\bibinfo {year} {2022})}\BibitemShut {NoStop}%
\bibitem [{\citenamefont {Tanzilli}\ \emph {et~al.}(2005)\citenamefont {Tanzilli}, \citenamefont {Tittel}, \citenamefont {Halder} \emph {et~al.}}]{12}%
  \BibitemOpen
  \bibfield  {author} {\bibinfo {author} {\bibfnamefont {S.}~\bibnamefont {Tanzilli}}, \bibinfo {author} {\bibfnamefont {W.}~\bibnamefont {Tittel}}, \bibinfo {author} {\bibfnamefont {M.}~\bibnamefont {Halder}}, \emph {et~al.},\ }\bibfield  {title} {\bibinfo {title} {A photonic quantum information interface},\ }\href {https://doi.org/10.1038/nature04009} {\bibfield  {journal} {\bibinfo  {journal} {Nature}\ }\textbf {\bibinfo {volume} {437}},\ \bibinfo {pages} {116} (\bibinfo {year} {2005})}\BibitemShut {NoStop}%
\bibitem [{\citenamefont {Ikuta}\ \emph {et~al.}(2018)\citenamefont {Ikuta}, \citenamefont {Kobayashi}, \citenamefont {Kawakami} \emph {et~al.}}]{13}%
  \BibitemOpen
  \bibfield  {author} {\bibinfo {author} {\bibfnamefont {R.}~\bibnamefont {Ikuta}}, \bibinfo {author} {\bibfnamefont {T.}~\bibnamefont {Kobayashi}}, \bibinfo {author} {\bibfnamefont {T.}~\bibnamefont {Kawakami}}, \emph {et~al.},\ }\bibfield  {title} {\bibinfo {title} {Polarization insensitive frequency conversion for an atom-photon entanglement distribution via a telecom network},\ }\href {https://doi.org/10.1038/s41467-018-04338-x} {\bibfield  {journal} {\bibinfo  {journal} {Nat. Commun.}\ }\textbf {\bibinfo {volume} {9}},\ \bibinfo {pages} {1997} (\bibinfo {year} {2018})}\BibitemShut {NoStop}%
\bibitem [{\citenamefont {Fisher}\ \emph {et~al.}(2021)\citenamefont {Fisher}, \citenamefont {Cernansky}, \citenamefont {Haylock} \emph {et~al.}}]{9}%
  \BibitemOpen
  \bibfield  {author} {\bibinfo {author} {\bibfnamefont {P.}~\bibnamefont {Fisher}}, \bibinfo {author} {\bibfnamefont {R.}~\bibnamefont {Cernansky}}, \bibinfo {author} {\bibfnamefont {B.}~\bibnamefont {Haylock}}, \emph {et~al.},\ }\bibfield  {title} {\bibinfo {title} {Single photon frequency conversion for frequency multiplexed quantum networks in the telecom band},\ }\href {https://doi.org/10.1103/PhysRevLett.127.023602} {\bibfield  {journal} {\bibinfo  {journal} {Phys. Rev. Lett.}\ }\textbf {\bibinfo {volume} {127}},\ \bibinfo {pages} {023602} (\bibinfo {year} {2021})}\BibitemShut {NoStop}%
\bibitem [{\citenamefont {Luo}\ \emph {et~al.}(2022)\citenamefont {Luo}, \citenamefont {Yu}, \citenamefont {Liu} \emph {et~al.}}]{11}%
  \BibitemOpen
  \bibfield  {author} {\bibinfo {author} {\bibfnamefont {X.~Y.}\ \bibnamefont {Luo}}, \bibinfo {author} {\bibfnamefont {Y.}~\bibnamefont {Yu}}, \bibinfo {author} {\bibfnamefont {J.~L.}\ \bibnamefont {Liu}}, \emph {et~al.},\ }\bibfield  {title} {\bibinfo {title} {Postselected entanglement between two atomic ensembles separated by 12.5 km},\ }\href {https://doi.org/10.1103/PhysRevLett.129.050503} {\bibfield  {journal} {\bibinfo  {journal} {Phys. Rev. Lett.}\ }\textbf {\bibinfo {volume} {129}},\ \bibinfo {pages} {050503} (\bibinfo {year} {2022})}\BibitemShut {NoStop}%
\bibitem [{\citenamefont {Van~Leent}\ \emph {et~al.}(2022)\citenamefont {Van~Leent}, \citenamefont {Bock}, \citenamefont {Fertig}, \citenamefont {Garthoff}, \citenamefont {Eppelt}, \citenamefont {Zhou}, \citenamefont {Malik}, \citenamefont {Seubert}, \citenamefont {Bauer}, \citenamefont {Rosenfeld} \emph {et~al.}}]{van2022entangling}%
  \BibitemOpen
  \bibfield  {author} {\bibinfo {author} {\bibfnamefont {T.}~\bibnamefont {Van~Leent}}, \bibinfo {author} {\bibfnamefont {M.}~\bibnamefont {Bock}}, \bibinfo {author} {\bibfnamefont {F.}~\bibnamefont {Fertig}}, \bibinfo {author} {\bibfnamefont {R.}~\bibnamefont {Garthoff}}, \bibinfo {author} {\bibfnamefont {S.}~\bibnamefont {Eppelt}}, \bibinfo {author} {\bibfnamefont {Y.}~\bibnamefont {Zhou}}, \bibinfo {author} {\bibfnamefont {P.}~\bibnamefont {Malik}}, \bibinfo {author} {\bibfnamefont {M.}~\bibnamefont {Seubert}}, \bibinfo {author} {\bibfnamefont {T.}~\bibnamefont {Bauer}}, \bibinfo {author} {\bibfnamefont {W.}~\bibnamefont {Rosenfeld}}, \emph {et~al.},\ }\bibfield  {title} {\bibinfo {title} {Entangling single atoms over 33 km telecom fibre},\ }\href@noop {} {\bibfield  {journal} {\bibinfo  {journal} {Nature}\ }\textbf {\bibinfo {volume} {607}},\ \bibinfo {pages} {69} (\bibinfo {year} {2022})}\BibitemShut {NoStop}%
\bibitem [{\citenamefont {Liu}\ \emph {et~al.}(2024)\citenamefont {Liu}, \citenamefont {Luo}, \citenamefont {Yu}, \citenamefont {Wang}, \citenamefont {Wang}, \citenamefont {Hu}, \citenamefont {Li}, \citenamefont {Zheng}, \citenamefont {Yao}, \citenamefont {Yan} \emph {et~al.}}]{liu2024creation}%
  \BibitemOpen
  \bibfield  {author} {\bibinfo {author} {\bibfnamefont {J.~L.}\ \bibnamefont {Liu}}, \bibinfo {author} {\bibfnamefont {X.~Y.}\ \bibnamefont {Luo}}, \bibinfo {author} {\bibfnamefont {Y.}~\bibnamefont {Yu}}, \bibinfo {author} {\bibfnamefont {C.-Y.}\ \bibnamefont {Wang}}, \bibinfo {author} {\bibfnamefont {B.}~\bibnamefont {Wang}}, \bibinfo {author} {\bibfnamefont {Y.}~\bibnamefont {Hu}}, \bibinfo {author} {\bibfnamefont {J.}~\bibnamefont {Li}}, \bibinfo {author} {\bibfnamefont {M.-Y.}\ \bibnamefont {Zheng}}, \bibinfo {author} {\bibfnamefont {B.}~\bibnamefont {Yao}}, \bibinfo {author} {\bibfnamefont {Z.}~\bibnamefont {Yan}}, \emph {et~al.},\ }\bibfield  {title} {\bibinfo {title} {Creation of memory--memory entanglement in a metropolitan quantum network},\ }\href@noop {} {\bibfield  {journal} {\bibinfo  {journal} {Nature}\ }\textbf {\bibinfo {volume} {629}},\ \bibinfo {pages} {579} (\bibinfo {year} {2024})}\BibitemShut {NoStop}%
\bibitem [{\citenamefont {Sidhu}\ \emph {et~al.}(2021)\citenamefont {Sidhu}, \citenamefont {Joshi}, \citenamefont {Gündoğan} \emph {et~al.}}]{16}%
  \BibitemOpen
  \bibfield  {author} {\bibinfo {author} {\bibfnamefont {J.~S.}\ \bibnamefont {Sidhu}}, \bibinfo {author} {\bibfnamefont {S.~K.}\ \bibnamefont {Joshi}}, \bibinfo {author} {\bibfnamefont {M.}~\bibnamefont {Gündoğan}}, \emph {et~al.},\ }\bibfield  {title} {\bibinfo {title} {Advances in space quantum communications},\ }\href {https://doi.org/10.1049/qtc2.12015} {\bibfield  {journal} {\bibinfo  {journal} {IET Quantum Commun.}\ }\textbf {\bibinfo {volume} {2}},\ \bibinfo {pages} {182} (\bibinfo {year} {2021})}\BibitemShut {NoStop}%
\bibitem [{\citenamefont {Liao}\ \emph {et~al.}(2017)\citenamefont {Liao}, \citenamefont {Yong}, \citenamefont {Liu} \emph {et~al.}}]{17}%
  \BibitemOpen
  \bibfield  {author} {\bibinfo {author} {\bibfnamefont {S.~K.}\ \bibnamefont {Liao}}, \bibinfo {author} {\bibfnamefont {H.~L.}\ \bibnamefont {Yong}}, \bibinfo {author} {\bibfnamefont {C.}~\bibnamefont {Liu}}, \emph {et~al.},\ }\bibfield  {title} {\bibinfo {title} {Long-distance free-space quantum key distribution in daylight towards inter-satellite communication},\ }\href {https://doi.org/10.1038/nphoton.2017.116} {\bibfield  {journal} {\bibinfo  {journal} {Nat. Photon.}\ }\textbf {\bibinfo {volume} {11}},\ \bibinfo {pages} {509} (\bibinfo {year} {2017})}\BibitemShut {NoStop}%
\bibitem [{\citenamefont {Shangguan}\ \emph {et~al.}(2016)\citenamefont {Shangguan}, \citenamefont {Xia}, \citenamefont {Wang} \emph {et~al.}}]{18}%
  \BibitemOpen
  \bibfield  {author} {\bibinfo {author} {\bibfnamefont {M.~J.}\ \bibnamefont {Shangguan}}, \bibinfo {author} {\bibfnamefont {H.~Y.}\ \bibnamefont {Xia}}, \bibinfo {author} {\bibfnamefont {C.}~\bibnamefont {Wang}}, \emph {et~al.},\ }\bibfield  {title} {\bibinfo {title} {All-fiber upconversion high spectral resolution wind lidar using a fabry-perot interferometer},\ }\href {https://doi.org/10.1364/OE.24.019322} {\bibfield  {journal} {\bibinfo  {journal} {Opt. Express}\ }\textbf {\bibinfo {volume} {24}},\ \bibinfo {pages} {19322} (\bibinfo {year} {2016})}\BibitemShut {NoStop}%
\bibitem [{\citenamefont {Wang}\ \emph {et~al.}(2021)\citenamefont {Wang}, \citenamefont {Zheng}, \citenamefont {Han}, \citenamefont {Huang}, \citenamefont {Xie}, \citenamefont {Xu}, \citenamefont {Zhang},\ and\ \citenamefont {Pan}}]{wang2021non}%
  \BibitemOpen
  \bibfield  {author} {\bibinfo {author} {\bibfnamefont {B.}~\bibnamefont {Wang}}, \bibinfo {author} {\bibfnamefont {M.~Y.}\ \bibnamefont {Zheng}}, \bibinfo {author} {\bibfnamefont {J.~J.}\ \bibnamefont {Han}}, \bibinfo {author} {\bibfnamefont {X.}~\bibnamefont {Huang}}, \bibinfo {author} {\bibfnamefont {X.-P.}\ \bibnamefont {Xie}}, \bibinfo {author} {\bibfnamefont {F.}~\bibnamefont {Xu}}, \bibinfo {author} {\bibfnamefont {Q.}~\bibnamefont {Zhang}},\ and\ \bibinfo {author} {\bibfnamefont {J.-W.}\ \bibnamefont {Pan}},\ }\bibfield  {title} {\bibinfo {title} {Non-line-of-sight imaging with picosecond temporal resolution},\ }\href@noop {} {\bibfield  {journal} {\bibinfo  {journal} {Phys. Rev. Lett.}\ }\textbf {\bibinfo {volume} {127}},\ \bibinfo {pages} {053602} (\bibinfo {year} {2021})}\BibitemShut {NoStop}%
\bibitem [{\citenamefont {Wang}\ \emph {et~al.}(2023{\natexlab{a}})\citenamefont {Wang}, \citenamefont {Huang}, \citenamefont {Fang}, \citenamefont {Yan}, \citenamefont {Wu},\ and\ \citenamefont {Zeng}}]{wang2023mid}%
  \BibitemOpen
  \bibfield  {author} {\bibinfo {author} {\bibfnamefont {Y.}~\bibnamefont {Wang}}, \bibinfo {author} {\bibfnamefont {K.}~\bibnamefont {Huang}}, \bibinfo {author} {\bibfnamefont {J.}~\bibnamefont {Fang}}, \bibinfo {author} {\bibfnamefont {M.}~\bibnamefont {Yan}}, \bibinfo {author} {\bibfnamefont {E.}~\bibnamefont {Wu}},\ and\ \bibinfo {author} {\bibfnamefont {H.}~\bibnamefont {Zeng}},\ }\bibfield  {title} {\bibinfo {title} {Mid-infrared single-pixel imaging at the single-photon level},\ }\href@noop {} {\bibfield  {journal} {\bibinfo  {journal} {Nat. Commun.}\ }\textbf {\bibinfo {volume} {14}},\ \bibinfo {pages} {1073} (\bibinfo {year} {2023}{\natexlab{a}})}\BibitemShut {NoStop}%
\bibitem [{\citenamefont {Takesue}(2008)}]{20}%
  \BibitemOpen
  \bibfield  {author} {\bibinfo {author} {\bibfnamefont {H.}~\bibnamefont {Takesue}},\ }\bibfield  {title} {\bibinfo {title} {Erasing distinguishability using quantum frequency up-conversion},\ }\href {https://doi.org/10.1103/PhysRevLett.101.173901} {\bibfield  {journal} {\bibinfo  {journal} {Phys. Rev. Lett.}\ }\textbf {\bibinfo {volume} {101}},\ \bibinfo {pages} {173901} (\bibinfo {year} {2008})}\BibitemShut {NoStop}%
\bibitem [{\citenamefont {Darr{\'e}}\ \emph {et~al.}(2016)\citenamefont {Darr{\'e}}, \citenamefont {Baudoin}, \citenamefont {Gomes}, \citenamefont {Scott}, \citenamefont {Delage}, \citenamefont {Grossard}, \citenamefont {Sturmann}, \citenamefont {Farrington}, \citenamefont {Reynaud},\ and\ \citenamefont {Brummelaar}}]{darre2016first}%
  \BibitemOpen
  \bibfield  {author} {\bibinfo {author} {\bibfnamefont {P.}~\bibnamefont {Darr{\'e}}}, \bibinfo {author} {\bibfnamefont {R.}~\bibnamefont {Baudoin}}, \bibinfo {author} {\bibfnamefont {J.-T.}\ \bibnamefont {Gomes}}, \bibinfo {author} {\bibfnamefont {N.}~\bibnamefont {Scott}}, \bibinfo {author} {\bibfnamefont {L.}~\bibnamefont {Delage}}, \bibinfo {author} {\bibfnamefont {L.}~\bibnamefont {Grossard}}, \bibinfo {author} {\bibfnamefont {J.}~\bibnamefont {Sturmann}}, \bibinfo {author} {\bibfnamefont {C.}~\bibnamefont {Farrington}}, \bibinfo {author} {\bibfnamefont {F.}~\bibnamefont {Reynaud}},\ and\ \bibinfo {author} {\bibfnamefont {T.~T.}\ \bibnamefont {Brummelaar}},\ }\bibfield  {title} {\bibinfo {title} {First on-sky fringes with an up-conversion interferometer tested on a telescope array},\ }\href@noop {} {\bibfield  {journal} {\bibinfo  {journal} {Phys. Rev. Lett.}\ }\textbf {\bibinfo {volume} {117}},\ \bibinfo {pages} {233902} (\bibinfo {year} {2016})}\BibitemShut {NoStop}%
\bibitem [{\citenamefont {Qu}\ \emph {et~al.}(2019)\citenamefont {Qu}, \citenamefont {Cotler}, \citenamefont {Ma} \emph {et~al.}}]{21}%
  \BibitemOpen
  \bibfield  {author} {\bibinfo {author} {\bibfnamefont {L.~Y.}\ \bibnamefont {Qu}}, \bibinfo {author} {\bibfnamefont {J.}~\bibnamefont {Cotler}}, \bibinfo {author} {\bibfnamefont {F.}~\bibnamefont {Ma}}, \emph {et~al.},\ }\bibfield  {title} {\bibinfo {title} {Color erasure detectors enable chromatic interferometry},\ }\href {https://doi.org/10.1103/PhysRevLett.123.243601} {\bibfield  {journal} {\bibinfo  {journal} {Phys. Rev. Lett.}\ }\textbf {\bibinfo {volume} {123}},\ \bibinfo {pages} {243601} (\bibinfo {year} {2019})}\BibitemShut {NoStop}%
\bibitem [{\citenamefont {Liu}\ \emph {et~al.}(2021)\citenamefont {Liu}, \citenamefont {Qu}, \citenamefont {Wu}, \citenamefont {Cotler}, \citenamefont {Ma}, \citenamefont {Zheng}, \citenamefont {Xie}, \citenamefont {Chen}, \citenamefont {Zhang}, \citenamefont {Wilczek} \emph {et~al.}}]{liu2021improved}%
  \BibitemOpen
  \bibfield  {author} {\bibinfo {author} {\bibfnamefont {L.~C.}\ \bibnamefont {Liu}}, \bibinfo {author} {\bibfnamefont {L.~Y.}\ \bibnamefont {Qu}}, \bibinfo {author} {\bibfnamefont {C.}~\bibnamefont {Wu}}, \bibinfo {author} {\bibfnamefont {J.}~\bibnamefont {Cotler}}, \bibinfo {author} {\bibfnamefont {F.}~\bibnamefont {Ma}}, \bibinfo {author} {\bibfnamefont {M.-Y.}\ \bibnamefont {Zheng}}, \bibinfo {author} {\bibfnamefont {X.-P.}\ \bibnamefont {Xie}}, \bibinfo {author} {\bibfnamefont {Y.-A.}\ \bibnamefont {Chen}}, \bibinfo {author} {\bibfnamefont {Q.}~\bibnamefont {Zhang}}, \bibinfo {author} {\bibfnamefont {F.}~\bibnamefont {Wilczek}}, \emph {et~al.},\ }\bibfield  {title} {\bibinfo {title} {Improved spatial resolution achieved by chromatic intensity interferometry},\ }\href@noop {} {\bibfield  {journal} {\bibinfo  {journal} {Phys. Rev. Lett.}\ }\textbf {\bibinfo {volume} {127}},\ \bibinfo {pages} {103601} (\bibinfo {year} {2021})}\BibitemShut {NoStop}%
\bibitem [{\citenamefont {Widarsson}\ \emph {et~al.}(2022)\citenamefont {Widarsson}, \citenamefont {Henriksson}, \citenamefont {Barrett} \emph {et~al.}}]{19}%
  \BibitemOpen
  \bibfield  {author} {\bibinfo {author} {\bibfnamefont {M.}~\bibnamefont {Widarsson}}, \bibinfo {author} {\bibfnamefont {M.}~\bibnamefont {Henriksson}}, \bibinfo {author} {\bibfnamefont {L.}~\bibnamefont {Barrett}}, \emph {et~al.},\ }\bibfield  {title} {\bibinfo {title} {Room temperature photon-counting lidar at 3 $\mu$m},\ }\href {https://doi.org/10.1364/AO.443938} {\bibfield  {journal} {\bibinfo  {journal} {Appl. Opt.}\ }\textbf {\bibinfo {volume} {61}},\ \bibinfo {pages} {884} (\bibinfo {year} {2022})}\BibitemShut {NoStop}%
\bibitem [{\citenamefont {Zheng}\ \emph {et~al.}(2020)\citenamefont {Zheng}, \citenamefont {Yao}, \citenamefont {Wang}, \citenamefont {Xie}, \citenamefont {Zhang},\ and\ \citenamefont {Pan}}]{PhysRevApplied.14.034035}%
  \BibitemOpen
  \bibfield  {author} {\bibinfo {author} {\bibfnamefont {M.~Y.}\ \bibnamefont {Zheng}}, \bibinfo {author} {\bibfnamefont {Q.}~\bibnamefont {Yao}}, \bibinfo {author} {\bibfnamefont {B.}~\bibnamefont {Wang}}, \bibinfo {author} {\bibfnamefont {X.-P.}\ \bibnamefont {Xie}}, \bibinfo {author} {\bibfnamefont {Q.}~\bibnamefont {Zhang}},\ and\ \bibinfo {author} {\bibfnamefont {J.-W.}\ \bibnamefont {Pan}},\ }\bibfield  {title} {\bibinfo {title} {Integrated multichannel lithium niobate waveguides for quantum frequency conversion},\ }\href@noop {} {\bibfield  {journal} {\bibinfo  {journal} {Phys. Rev. Appl.}\ }\textbf {\bibinfo {volume} {14}},\ \bibinfo {pages} {034035} (\bibinfo {year} {2020})}\BibitemShut {NoStop}%
\bibitem [{\citenamefont {Zhu}\ \emph {et~al.}(2021)\citenamefont {Zhu}, \citenamefont {Shao}, \citenamefont {Yu} \emph {et~al.}}]{25}%
  \BibitemOpen
  \bibfield  {author} {\bibinfo {author} {\bibfnamefont {D.}~\bibnamefont {Zhu}}, \bibinfo {author} {\bibfnamefont {L.}~\bibnamefont {Shao}}, \bibinfo {author} {\bibfnamefont {M.}~\bibnamefont {Yu}}, \emph {et~al.},\ }\bibfield  {title} {\bibinfo {title} {Integrated photonics on thin-film lithium niobate},\ }\href {https://doi.org/10.1364/AOP.411024} {\bibfield  {journal} {\bibinfo  {journal} {Adv. Opt. Photon.}\ }\textbf {\bibinfo {volume} {13}},\ \bibinfo {pages} {242} (\bibinfo {year} {2021})}\BibitemShut {NoStop}%
\bibitem [{\citenamefont {Boes}\ \emph {et~al.}(2023)\citenamefont {Boes} \emph {et~al.}}]{26}%
  \BibitemOpen
  \bibfield  {author} {\bibinfo {author} {\bibfnamefont {A.}~\bibnamefont {Boes}} \emph {et~al.},\ }\bibfield  {title} {\bibinfo {title} {Lithium niobate photonics: unlocking the electromagnetic spectrum},\ }\href {https://doi.org/10.1126/science.abj4396} {\bibfield  {journal} {\bibinfo  {journal} {Science}\ }\textbf {\bibinfo {volume} {379}},\ \bibinfo {pages} {eabj4396} (\bibinfo {year} {2023})}\BibitemShut {NoStop}%
\bibitem [{\citenamefont {Niu}\ \emph {et~al.}(2020)\citenamefont {Niu}, \citenamefont {Lin}, \citenamefont {Liu}, \citenamefont {Chen}, \citenamefont {Hu}, \citenamefont {Zhang}, \citenamefont {Cai}, \citenamefont {Gong}, \citenamefont {Xie},\ and\ \citenamefont {Zhu}}]{niu2020optimizing}%
  \BibitemOpen
  \bibfield  {author} {\bibinfo {author} {\bibfnamefont {Y.}~\bibnamefont {Niu}}, \bibinfo {author} {\bibfnamefont {C.}~\bibnamefont {Lin}}, \bibinfo {author} {\bibfnamefont {X.}~\bibnamefont {Liu}}, \bibinfo {author} {\bibfnamefont {Y.}~\bibnamefont {Chen}}, \bibinfo {author} {\bibfnamefont {X.}~\bibnamefont {Hu}}, \bibinfo {author} {\bibfnamefont {Y.}~\bibnamefont {Zhang}}, \bibinfo {author} {\bibfnamefont {X.}~\bibnamefont {Cai}}, \bibinfo {author} {\bibfnamefont {Y.-X.}\ \bibnamefont {Gong}}, \bibinfo {author} {\bibfnamefont {Z.}~\bibnamefont {Xie}},\ and\ \bibinfo {author} {\bibfnamefont {S.}~\bibnamefont {Zhu}},\ }\bibfield  {title} {\bibinfo {title} {Optimizing the efficiency of a periodically poled {LNOI} waveguide using in situ monitoring of the ferroelectric domains},\ }\href@noop {} {\bibfield  {journal} {\bibinfo  {journal} {Appl. Phys. Lett.}\ }\textbf {\bibinfo {volume} {116}},\ \bibinfo {pages} {101104} (\bibinfo {year} {2020})}\BibitemShut {NoStop}%
\bibitem [{\citenamefont {Fan}\ \emph {et~al.}(2021)\citenamefont {Fan}, \citenamefont {Ma}, \citenamefont {Chen} \emph {et~al.}}]{27}%
  \BibitemOpen
  \bibfield  {author} {\bibinfo {author} {\bibfnamefont {H.}~\bibnamefont {Fan}}, \bibinfo {author} {\bibfnamefont {Z.~H.}\ \bibnamefont {Ma}}, \bibinfo {author} {\bibfnamefont {J.~Y.}\ \bibnamefont {Chen}}, \emph {et~al.},\ }\bibfield  {title} {\bibinfo {title} {Photon conversion in thin-film lithium niobate nanowaveguides: a noise analysis},\ }\href {https://doi.org/10.1364/JOSAB.426488} {\bibfield  {journal} {\bibinfo  {journal} {J. Opt. Soc. Am. B}\ }\textbf {\bibinfo {volume} {38}},\ \bibinfo {pages} {2172} (\bibinfo {year} {2021})}\BibitemShut {NoStop}%
\bibitem [{\citenamefont {Wang}\ \emph {et~al.}(2023{\natexlab{b}})\citenamefont {Wang}, \citenamefont {Jiao}, \citenamefont {Wang} \emph {et~al.}}]{29}%
  \BibitemOpen
  \bibfield  {author} {\bibinfo {author} {\bibfnamefont {X.~N.}\ \bibnamefont {Wang}}, \bibinfo {author} {\bibfnamefont {X.~F.}\ \bibnamefont {Jiao}}, \bibinfo {author} {\bibfnamefont {B.}~\bibnamefont {Wang}}, \emph {et~al.},\ }\bibfield  {title} {\bibinfo {title} {Quantum frequency conversion and single-photon detection with lithium niobate nanophotonic chips},\ }\href {https://doi.org/10.1038/s41534-023-00704-w} {\bibfield  {journal} {\bibinfo  {journal} {npj Quantum Inf.}\ }\textbf {\bibinfo {volume} {9}},\ \bibinfo {pages} {38} (\bibinfo {year} {2023}{\natexlab{b}})}\BibitemShut {NoStop}%
\bibitem [{\citenamefont {Chen}\ \emph {et~al.}(2021)\citenamefont {Chen}, \citenamefont {Li}, \citenamefont {Ma} \emph {et~al.}}]{28}%
  \BibitemOpen
  \bibfield  {author} {\bibinfo {author} {\bibfnamefont {J.~Y.}\ \bibnamefont {Chen}}, \bibinfo {author} {\bibfnamefont {Z.}~\bibnamefont {Li}}, \bibinfo {author} {\bibfnamefont {Z.~H.}\ \bibnamefont {Ma}}, \emph {et~al.},\ }\bibfield  {title} {\bibinfo {title} {Photon conversion and interaction in a quasi-phase-matched microresonator},\ }\href {https://doi.org/10.1103/PhysRevApplied.16.064004} {\bibfield  {journal} {\bibinfo  {journal} {Phys. Rev. Appl.}\ }\textbf {\bibinfo {volume} {16}},\ \bibinfo {pages} {064004} (\bibinfo {year} {2021})}\BibitemShut {NoStop}%
\bibitem [{\citenamefont {Breunig}(2016)}]{30}%
  \BibitemOpen
  \bibfield  {author} {\bibinfo {author} {\bibfnamefont {I.}~\bibnamefont {Breunig}},\ }\bibfield  {title} {\bibinfo {title} {Three-wave mixing in whispering gallery resonators},\ }\href {https://doi.org/10.1002/lpor.201500252} {\bibfield  {journal} {\bibinfo  {journal} {Laser Photonics Rev.}\ }\textbf {\bibinfo {volume} {10}},\ \bibinfo {pages} {569} (\bibinfo {year} {2016})}\BibitemShut {NoStop}%
\bibitem [{\citenamefont {Lu}\ \emph {et~al.}(2019)\citenamefont {Lu}, \citenamefont {Surya}, \citenamefont {Liu} \emph {et~al.}}]{31}%
  \BibitemOpen
  \bibfield  {author} {\bibinfo {author} {\bibfnamefont {J.~J.}\ \bibnamefont {Lu}}, \bibinfo {author} {\bibfnamefont {J.~B.}\ \bibnamefont {Surya}}, \bibinfo {author} {\bibfnamefont {X.~W.}\ \bibnamefont {Liu}}, \emph {et~al.},\ }\bibfield  {title} {\bibinfo {title} {Periodically poled thin-film lithium niobate microring resonators with a second-harmonic generation efficiency of 250,000\%/{W}},\ }\href {https://doi.org/10.1364/OPTICA.6.001455} {\bibfield  {journal} {\bibinfo  {journal} {Optica}\ }\textbf {\bibinfo {volume} {6}},\ \bibinfo {pages} {1455} (\bibinfo {year} {2019})}\BibitemShut {NoStop}%
\bibitem [{\citenamefont {Jiang}\ \emph {et~al.}(2018)\citenamefont {Jiang}, \citenamefont {Wu},\ and\ \citenamefont {Dai}}]{32}%
  \BibitemOpen
  \bibfield  {author} {\bibinfo {author} {\bibfnamefont {X.~H.}\ \bibnamefont {Jiang}}, \bibinfo {author} {\bibfnamefont {H.}~\bibnamefont {Wu}},\ and\ \bibinfo {author} {\bibfnamefont {D.~X.}\ \bibnamefont {Dai}},\ }\bibfield  {title} {\bibinfo {title} {Low-loss and low-crosstalk multimode waveguide bend on silicon},\ }\href {https://doi.org/10.1364/OE.26.017680} {\bibfield  {journal} {\bibinfo  {journal} {Opt. Express}\ }\textbf {\bibinfo {volume} {26}},\ \bibinfo {pages} {17680} (\bibinfo {year} {2018})}\BibitemShut {NoStop}%
\bibitem [{\citenamefont {Yu}\ \emph {et~al.}(2022)\citenamefont {Yu}, \citenamefont {Barton~III}, \citenamefont {Cheng}, \citenamefont {Reimer}, \citenamefont {Kharel}, \citenamefont {He}, \citenamefont {Shao}, \citenamefont {Zhu}, \citenamefont {Hu}, \citenamefont {Grant} \emph {et~al.}}]{ref27}%
  \BibitemOpen
  \bibfield  {author} {\bibinfo {author} {\bibfnamefont {M.}~\bibnamefont {Yu}}, \bibinfo {author} {\bibfnamefont {D.}~\bibnamefont {Barton~III}}, \bibinfo {author} {\bibfnamefont {R.}~\bibnamefont {Cheng}}, \bibinfo {author} {\bibfnamefont {C.}~\bibnamefont {Reimer}}, \bibinfo {author} {\bibfnamefont {P.}~\bibnamefont {Kharel}}, \bibinfo {author} {\bibfnamefont {L.}~\bibnamefont {He}}, \bibinfo {author} {\bibfnamefont {L.}~\bibnamefont {Shao}}, \bibinfo {author} {\bibfnamefont {D.}~\bibnamefont {Zhu}}, \bibinfo {author} {\bibfnamefont {Y.}~\bibnamefont {Hu}}, \bibinfo {author} {\bibfnamefont {H.~R.}\ \bibnamefont {Grant}}, \emph {et~al.},\ }\bibfield  {title} {\bibinfo {title} {Integrated femtosecond pulse generator on thin-film lithium niobate},\ }\href@noop {} {\bibfield  {journal} {\bibinfo  {journal} {Nature}\ }\textbf {\bibinfo {volume} {612}},\ \bibinfo {pages} {252} (\bibinfo {year} {2022})}\BibitemShut {NoStop}%
\bibitem [{\citenamefont {Ling}\ \emph {et~al.}(2024)\citenamefont {Ling}, \citenamefont {Gao}, \citenamefont {Xue}, \citenamefont {Hu}, \citenamefont {Li}, \citenamefont {Zhang}, \citenamefont {Javid}, \citenamefont {Lopez-Rios}, \citenamefont {Staffa},\ and\ \citenamefont {Lin}}]{ref26}%
  \BibitemOpen
  \bibfield  {author} {\bibinfo {author} {\bibfnamefont {J.}~\bibnamefont {Ling}}, \bibinfo {author} {\bibfnamefont {Z.}~\bibnamefont {Gao}}, \bibinfo {author} {\bibfnamefont {S.}~\bibnamefont {Xue}}, \bibinfo {author} {\bibfnamefont {Q.}~\bibnamefont {Hu}}, \bibinfo {author} {\bibfnamefont {M.}~\bibnamefont {Li}}, \bibinfo {author} {\bibfnamefont {K.}~\bibnamefont {Zhang}}, \bibinfo {author} {\bibfnamefont {U.~A.}\ \bibnamefont {Javid}}, \bibinfo {author} {\bibfnamefont {R.}~\bibnamefont {Lopez-Rios}}, \bibinfo {author} {\bibfnamefont {J.}~\bibnamefont {Staffa}},\ and\ \bibinfo {author} {\bibfnamefont {Q.}~\bibnamefont {Lin}},\ }\bibfield  {title} {\bibinfo {title} {Electrically empowered microcomb laser},\ }\href@noop {} {\bibfield  {journal} {\bibinfo  {journal} {Nat. Commun.}\ }\textbf {\bibinfo {volume} {15}},\ \bibinfo {pages} {4192} (\bibinfo {year} {2024})}\BibitemShut {NoStop}%
\bibitem [{\citenamefont {Zhang}\ \emph {et~al.}(2025)\citenamefont {Zhang}, \citenamefont {Wang}, \citenamefont {Denney}, \citenamefont {Riemensberger}, \citenamefont {Lihachev}, \citenamefont {Hu}, \citenamefont {Kao}, \citenamefont {Bl{\'e}sin}, \citenamefont {Kuznetsov}, \citenamefont {Li} \emph {et~al.}}]{ref28}%
  \BibitemOpen
  \bibfield  {author} {\bibinfo {author} {\bibfnamefont {J.}~\bibnamefont {Zhang}}, \bibinfo {author} {\bibfnamefont {C.}~\bibnamefont {Wang}}, \bibinfo {author} {\bibfnamefont {C.}~\bibnamefont {Denney}}, \bibinfo {author} {\bibfnamefont {J.}~\bibnamefont {Riemensberger}}, \bibinfo {author} {\bibfnamefont {G.}~\bibnamefont {Lihachev}}, \bibinfo {author} {\bibfnamefont {J.}~\bibnamefont {Hu}}, \bibinfo {author} {\bibfnamefont {W.}~\bibnamefont {Kao}}, \bibinfo {author} {\bibfnamefont {T.}~\bibnamefont {Bl{\'e}sin}}, \bibinfo {author} {\bibfnamefont {N.}~\bibnamefont {Kuznetsov}}, \bibinfo {author} {\bibfnamefont {Z.}~\bibnamefont {Li}}, \emph {et~al.},\ }\bibfield  {title} {\bibinfo {title} {Ultrabroadband integrated electro-optic frequency comb in lithium tantalate},\ }\href@noop {} {\bibfield  {journal} {\bibinfo  {journal} {Nature}\ }\textbf {\bibinfo {volume} {637}},\ \bibinfo {pages} {1096–1103} (\bibinfo {year} {2025})}\BibitemShut {NoStop}%
\end{thebibliography}%

\newpage
\thispagestyle{empty}
\begin{center}
\vspace*{2cm}
{\LARGE\textbf{Supplementary Materials for: Electrically pumped ultra-efficient quantum frequency conversion on thin film lithium niobate chip
}}\\[1cm]
\end{center}
\vspace{2cm}

\section*{Supplementary Note 1: Transmission Simulations for Couplers}

Fig. 5 shows the electric field transmission simulations of the signal, pump, and SF at coupler A and coupler B. In coupler A (Fig. 5(a)--(c)), the signal couples into the microring with a strength of $\kappa_{\text{s,A}}^{2} \sim 0.03$ and a overlap of 98\% with the quasi-TM fundamental mode, ensuring dominant fundamental-mode excitation. The pump exhibits a similar coupling strength $\kappa_{\text{p,A}}^{2} \sim 0.03$ but a lower mode overlap of 71\%, exciting higher-order modes. This does not significantly affect the cavity enhancement and can be compensated by increasing pump power. The SF coupling in coupler A is suppressed ($\kappa_{\text{sf,A}}^{2} \sim 0.005$), minimizing its loss. Conversely, in coupler B (Fig. 5(d)--(f)), the SF couples out efficiently with a strength of $\kappa_{\text{sf,B}}^{2} \sim 0.05$ and a 96\% TM$_{00}$ overlap, while the signal and pump experience weak coupling ($\kappa_{\text{s,B}}^{2} \sim 0.004$ and $\kappa_{\text{p,B}}^{2} \sim 0.003$, respectively), preserving their intracavity energy.

\section*{Supplementary Note 2: Waveguide Taper Design}

\begin{figure}
\begin{center}
\begin{tabular}{c}
\includegraphics[height=6cm]{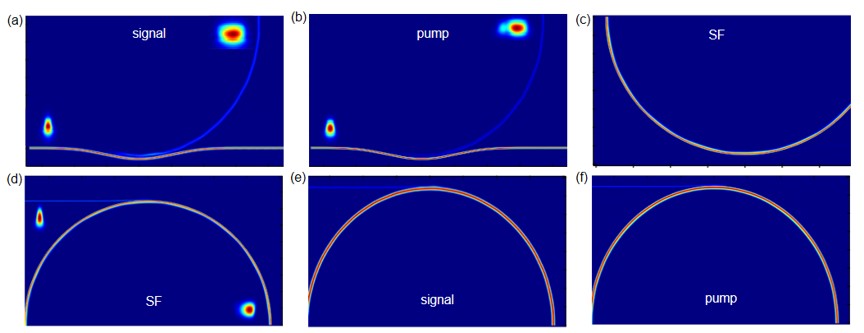}
\end{tabular}
\end{center}
\caption 
{\label{fig:example}
 Electric field transmission simulations of the signal, pump, and SF at (a)-(c) coupler A and (d)-(f) coupler B.} 
\end{figure} 

Fig. 6(a) shows the effective refractive indices of different waveguide widths. A slow and gradual broadening process leads to the formation of a hybridization region where partial coupling between the TM and TE modes occurs, resulting in hybrid modes. In this region, the TM mode gradually converts into the TE mode. Moreover, if the waveguide broadening is not sufficiently slow to maintain adiabatic evolution, mode hybridization can induce inter-mode crosstalk, causing multimode interference in subsequent transmission. To address this, we implemented an abrupt taper that rapidly expands the waveguide width, effectively bypassing the hybridization region while preserving the TM fundamental mode. In the gradual taper, where the waveguide width increases from 300 nm to 950 nm over a 300 $\mu$m length, the dominant mode gradually transitions from the TM mode to the TE mode. In contrast, the abrupt taper effectively bypasses the mode hybridization region, maintaining the TM-dominant mode despite some higher-order mode interference, as is show in Fig. 6(b). Fig. 6(c) shows the TM$_{00}$ mode transmission efficiency for different taper lengths.

\section*{Supplementary Note 3: Euler bend waveguide Design}

\begin{figure}
\begin{center}
\begin{tabular}{c}
\hspace*{-0.8cm}
\includegraphics[height=5cm]{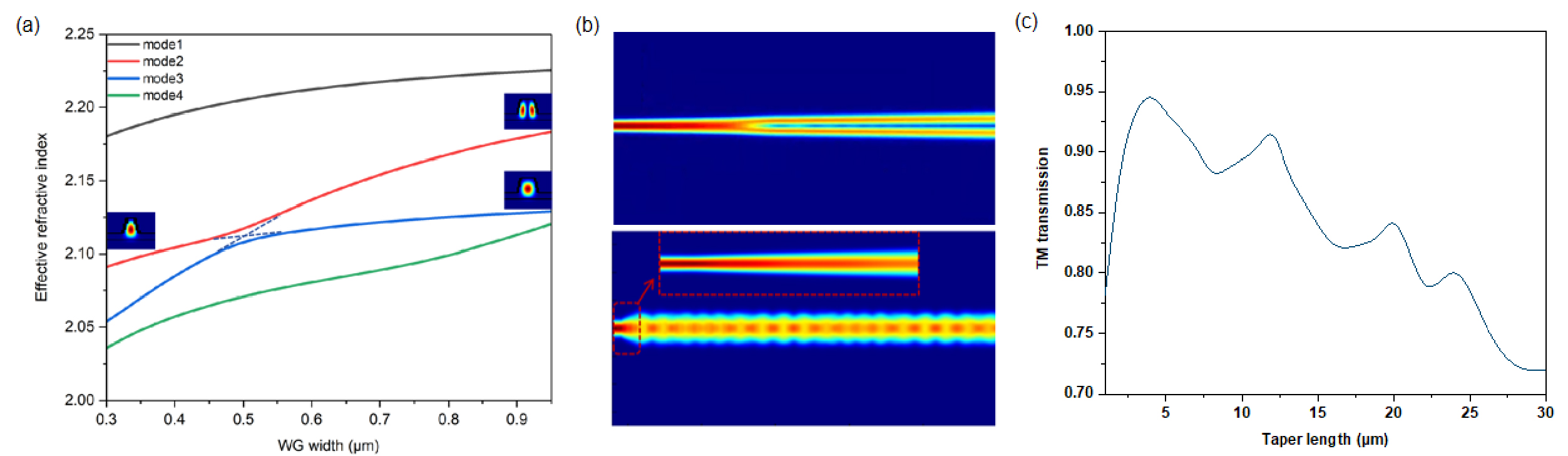}
\end{tabular}
\end{center}
\caption 
{\label{fig:example}
 (a) Effective refractive indices of different waveguide widths. (b) Electric field transmission of the adiabatic widening taper (top) over a 300 $\mu$m length and the abrupt taper (bottom). (c) $TM_{00}$ mode transmission efficiency for different taper lengths.} 
\end{figure} 
Then the SF transitions from a straight waveguide into a 950 nm-wide multimode bent waveguide, discontinuities in the mode field introduce excess loss and mode crosstalk, which can significantly impact on-chip integration and subsequent control. To mitigate these effects, we implemented a variable curvature Euler bent waveguide as shown in Fig. 7(a). Its curvature increases gradually from near zero along the curve. This smooth variation allows the mode field to seamlessly adapt its modal distribution during the transition from the straight waveguide to the bent waveguide, thereby enabling low-loss mode field transmission as shown in Fig. 7(b). The Euler curve is defined by its maximum and minimum curvature radii, and the effective radius is defined as the radius of a 90° arc covering the same region. For the SF wavelength of \(\sim\) 630 nm and a waveguide width of 950 nm, the optimized parameters are  : $R_{max} = 300$ $\mu \text{m} $,  $R_{min} = 28.5$ $\mu \text{m} $, and $R_{eff} = 50$ $\mu \text{m} $.
Simulation results  demonstrate that a 1/4 circular arc waveguide suffers from significant intermodal crosstalk, yielding approximately 20\% mode loss. In contrast, the Euler bent waveguide achieves a minimal mode loss of 0.25\%, ensuring efficient and stable optical field transmission. 

\begin{figure}
\begin{center}
\begin{tabular}{c}
\hspace*{-0.8cm}
\includegraphics[height=6cm]{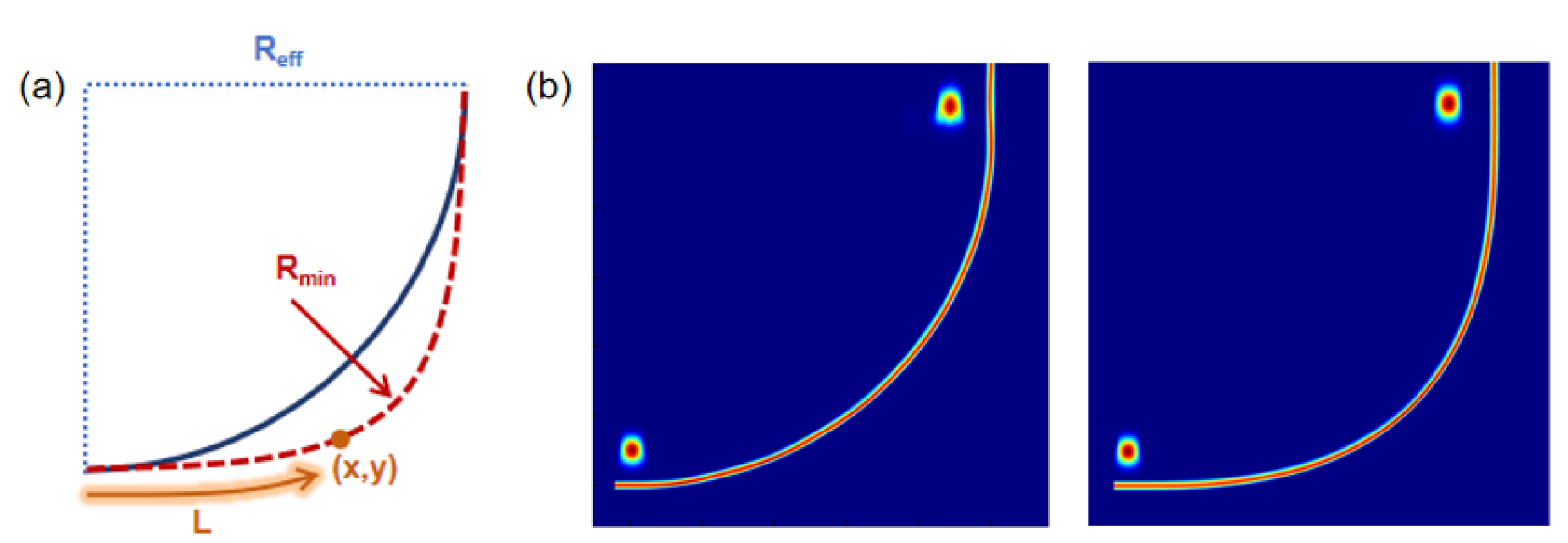}
\end{tabular}
\end{center}
\caption 
{\label{fig:example}
(a) Schematic of the 90°bent Euler curve (red dashed line) and arc (blue solid line). (b) Electric field transmission characteristics of SF in the 90°arc waveguide (left) and the Euler-bent waveguide (right). The optimized parameters of the Euler-bent waveguide includes $R_{\text{max}}$=300 $\mu$m, $R_{\text{min}}$=28.5 $\mu$m, $R_{\text{eff}}$=50 $\mu$m.  } 
\end{figure} 

\end{document}